\documentclass[journal]{IEEEtran}
\usepackage{soul}
\usepackage{color}

\usepackage{cite}

\ifCLASSINFOpdf
   \usepackage[pdftex]{graphicx}
\else
\fi
\usepackage{amsmath}
\usepackage{algorithm,algpseudocode}

\newcommand*\lowerb[1]{\underline{#1}}
\newcommand*\upperb[1]{\overline{#1}}

\DeclareMathOperator{\sign}{sign}
\DeclareMathOperator*{\argmin}{arg\,min}
\DeclareMathOperator{\diag}{blkdiag}

\begin{document}
%
\title{Distributed Maneuver Planning with Connected and Automated Vehicles for Boosting Traffic Efficiency}
%
%
%

\author{Nathan~Goulet,~\IEEEmembership{Student Member,~IEEE,}
        Beshah~Ayalew,~\IEEEmembership{Senior Member,~IEEE,}
\thanks{Nathan Goulet and Beshah Ayalew are with the Applied Dynamics \& Control Group at Clemson University - International Center for Automotive Research (CU-ICAR), 4 Research Dr., Greenville, SC 29607 USA
        {\tt\small \{ngoulet, beshah\}@clemson.edu}}
\thanks{This research was supported by an award from the U.S. Department of Energy Vehicle Technologies Office (Project No. DE-EE0008232)}}

\maketitle

\begin{abstract}
Connected and automated vehicles (CAVs) have the potential to improve traffic throughput and achieve a more efficient utilization of the available roadway infrastructure. They also have the potential to reduce energy consumption through traffic motion harmonization, even when operating in mixed traffic with other human-driven vehicles. The key to realizing these potentials are coordinated control schemes that can be implemented in a distributed manner with the CAVs. In this paper, we propose a distributed predictive control framework that features a two-dimensional maneuver planner incorporating explicit coordination constraints between connected vehicles operating in mixed traffic at various penetration levels. The framework includes a distributed implementation of a reference speed assigner that estimates local traffic speed from on-board measurements and communicated information. We present an extensive evaluation of the proposed framework in traffic micro-simulations at various CAV penetrations from traffic flow, energy use, and lane utilization points of view. Results are compared to a baseline scenario with no CAVs, as well as, a benchmark one-dimensional planner. 
\end{abstract}


%
\IEEEpeerreviewmaketitle

\vspace{-0.18in}
\section{Introduction}
%
%
%
%

\IEEEPARstart{D}{espite} recent improvements in vehicular fuel economy within the United States and across the globe \cite{USeia2019,GFEI2019}, the total amount of fuel consumed by the transportation industry continues to increase \cite{USeia2019,UNSD2019}. This is partially attributable to the fact that the number of vehicles and vehicle miles traveled (VMT) continues to increase \cite{USDOT2018,GFEI2019}. However, infrastructure development lags behind as both developed and emerging economies struggle to grow road networks commensurate with the transportation needs of their growing population \cite{Reed2019}. Left unchecked, this worsens traffic congestion, which exacerbates fuel use and pollution. This highlights the need to utilize the available infrastructure more efficiently in order to mitigate congestion and its negative economic and environmental impacts. Connected and automated vehicles (CAVs) have emerged as a potential technology to achieve this objective. 

An important deployment domain for CAVs is multi-lane interstates and highways, as a significant portion of VMT (over 30\%) occur on multi-lane highways \cite{USDOT2018}. CAVs are also expected to be first implemented in this domain as can be noted by currently available SAE Level 2 automated systems \cite{J3016202104}, such as General Motors' Super Cruise \cite{SuperCruise}. The multi-lane highway domain is one where CAVs can be more readily made to coordinate their maneuvers to boost traffic efficiency, in terms of both throughput and collective energy/fuel consumption.

Furthermore, in the foreseeable future, CAVs on multi-lane roads will be expected to operate in mixed traffic comprised of human-driven vehicles (HDVs) and other CAVs at various CAV penetration levels. As we review below, the potential of CAVs to improve traffic flow and energy efficiency in such mixed traffic has indeed received some attention. However, significant challenges remain in developing distributed maneuver planning algorithms for individual CAVs in a manner that balances egoistic goals with traffic harmonization objectives in mixed traffic. This paper proposes a distributed maneuver planning framework for CAVs operating in mixed traffic on multi-lane roads intended for use with vehicles equipped with an SAE level 3+ automated driving system (ADS) \cite{J3016202104}, however, with complements/modifications it may be utilized within SAE level 1 or higher ADS (e.g. as shown in \cite{Weiskircher2017} or later in Section \ref{sec:EvalSetup} here). Some components of our distributed planning and control framework previously appeared in our conference papers \cite{Goulet2019} and \cite{Goulet2020}. This paper makes the following expanded contributions:
\begin{itemize}
\item A more detailed and complete description of the distributed control  framework and methods used within;
\item A comprehensive analysis of  micro-simulations to demonstrate the potential improvements in traffic flow, travel time, and fuel consumption offered by the distributed control framework;
\item A comparison to a representative state-of-the-art cooperative adaptive cruise control (CACC) in order to isolate the benefits of a 2D maneuver planner over a 1D planner.
\end{itemize}

The paper is organized as follows: Section \ref{sec:RelatedWorks} reviews related works. Section \ref{sec:ControlFramework} presents the proposed distributed control and coordination framework and details its various computational components. Section \ref{sec:Results} presents results and discussions from evaluations of the framework implemented in extensive traffic micro-simulations. Section \ref{sec:Conclusion} presents concluding remarks.

\vspace{-0.15in}
\section{Related Works}\label{sec:RelatedWorks}

Early works in automated highway systems saw the potential of optimizing traffic flow on a lane level basis and sought to do this through optimal multi-agent speed and/or lane assignment \cite{Hall1999,Rao1994,Ramaswamy1997,Tsao1994}. Typically, the optimization problem was posed and solved in a decentralized manner by multiple higher level planners at the roadside. Therein, the planner would assign the optimal lane and/or travel speed to each vehicle within the link segment or neighborhood in their domain. A recent example of such a planner described in \cite{Jin2014} showed good potential for reducing fuel consumption and travel time. However, these types of approaches involve large investments on connected and computationally capable roadside infrastructure. An alternative and arguably more practical approach is to generate local maneuver plans for individual CAVs in a distributed yet coordinated manner by leveraging the vehicles' on-board computational and communication resources. Our review of related works in this section will focus on these latter approaches and a discussion of reported benefits of having such CAVs in traffic.

\vspace{-0.15in}
\subsection{Motion Planning and Control on Multi-Lane Roads}\label{sec:MotionPlanning}

There are two essential considerations in this topic: the coupling of the lateral control/lane selection and the longitudinal speed control of the individual vehicle, and the interactive coupling of individual vehicle control with those of other vehicles, i.e., the traffic control problem as a naturally multi-agent control problem. We begin by discussing some approaches to addressing the first consideration.

From an individual vehicle's control perspective, there are some established methods for motion planning that have been reviewed in detail in \cite{Claussmann2019}. Most of these focus on egoistic solutions, which undertake planning and re-planning of maneuvers for the ego vehicle as it advances in traffic. The recently popular ones can be grouped into path-finding algorithms and model predictive control.

Path-finding algorithms require two components: building a graph, and searching for a path within the graph. One realization of path-finding algorithms solves these two components separately: First, discretizing the state space in order to build a graph using methods such as sampling \cite{Geraerts2006}, or cell-decomposition \cite{Lavalle2006}; Second,  implementing a graph search algorithm, such as A* \cite{Hesse2010}, in order to find the optimal path within the graph. However, it can be computationally expensive to rebuild the graph at each call in dynamic environments.  Moreover, graph search methods are not typically formulated to incorporate lane decisions. The other realization of path-finding algorithms seeks to combine the two components. For example,  RRT samples the state-space and extends the graph using a dynamics model as part of the planning algorithm \cite{Kuwata2009}. The authors in \cite{Berntorp2016} and \cite{Berntorp2019} have expanded RRT to consider lane decisions. However, depending on the exploration method used, RRT can result in jerky behavior and re-checking nodes for collision during updates can become computationally costly.

Model predictive control (MPC) has emerged as an attractive method since it allows online optimization of multiple objectives and explicit consideration of constraints. In our previous work \cite{Weiskircher2017,Wang2015}, we have posed the individual vehicle guidance problem as a hierarchical control problem where lane decisions are separated from speed selections within a nonlinear model predictive control framework. In \cite{Wang2016} and \cite{Wang2019}, building from a hybrid dynamical systems view of the simultaneous planning of lane selection (discrete) and speed (continuous), the optimization problem is cast as a mixed integer problem (MIP) at each prediction horizon. Another example of a hierarchical MIP-based MPC formulation is presented in \cite{Kamal2016}. However, it is shown in \cite{Wang2019} that the computational complexity of the optimization problem to be solved for such MIP formulations grows as a high-order polynomial function of the number of lanes. Therefore, the decision variables were relaxed to be real values resulting in non-linear programs (NLP), where the computational complexity grows only as a low-order polynomial function of the number of lanes \cite{Wang2019}. A different approach to reducing the computational complexity of the MIP is proposed in \cite{Dollar2018}. This approach abstracts the lane change dynamics to a second order system driven by integer-indexed commanded lanes, so that the state dimensionality does not increase with an increase in the number of lanes. The imposed second order dynamics is, however, arbitrary and may not necessarily result in an optimal motion plan in practice.

One other important differentiating factor between the formulations used in \cite{Kamal2016} and \cite{Dollar2018}, and those used in \cite{Wang2016} and \cite{Wang2019}, is that the latter introduces a lane reference velocity. Therein, the lane decision considers information about the current traffic velocity within a given lane, going beyond simply minimizing accelerations. The idea of incorporating lane speed tracking cost was also proposed in \cite{WangM2015}. However, instead of blending the cost to track each lane, as done in \cite{Wang2019}, the lane decisions are discretized into a finite set in order to decompose the problem into a smaller set of sub-problems on which a sinusoidal lateral dynamics is imposed. Both assumptions may lead to sub-optimal trajectories. An alternative formulation that incorporates the current traffic state is proposed in \cite{Yoon2018}. This is accomplished by using the notion of social forces, including lane forces, within the objective function of the MPC. This approach was then implemented hierarchically \cite{Yoon2019} to aggregate lane loads on long horizons (using V2V and V2I information) to facilitate the planning computations.

It is possible to expand the above MPC-based approaches to a distributed or multi-agent implementation in order to account for the second consideration: the interaction between vehicles. Distributed MPC (DMPC) approaches still compute individual motion plans; that is, they are egoistic, allowing vehicles to have individual objectives/preferences. However, they take steps to account for the inherent coupling/interaction between the participating vehicles by leveraging communication \cite{Li2017,Scattolini2009}. There are two main categories of coordination schemes that accompany the distributed, and so decoupled, implementations sought with DMPC. The first, assumes the distributed agents coordinate their optimization iterations to achieve consensus on their shared variables before proceeding to the next MPC step \cite{DeOliveira2010, Negenborn2008}. This version, which draws techniques from parallelized distributed optimization such as augmented Lagrangian methods \cite{Bertsekas1997}, has a large communication overhead, but the solution could theoretically approach that of a centralized MPC solving for all vehicles. The second, assumes that computing agents communicate only after each agent completes the optimization for the next MPC step while the current local control actions are being applied; consequently, while the communication overhead is reduced, the available communicated information will have a one-step delay. Significant results have already been obtained with the latter DMPC approach. For example, this form of DMPC has been shown to stabilize vehicle formations by penalizing/constraining deviations from prior plans \cite{Dunbar2006,Bertrand2014}. Further by adding a terminal constraint and sufficient conditions on weights within the cost function, the stability, in a Lyapunov sense, is proven in \cite{Zheng2017} for various communication topologies. There are also DMPC formulations that account for interactions by adding an additional cost term in order to optimize not only the agents own cost, but also the cost incurred by their immediate neighbors \cite{WangM2015,WangM2016}.

Beyond DMPC, there are also control frameworks based on game theory and situation-aware (or interaction-aware) planning that inherently account for interactions between agents. Game theory based approaches, such as those in \cite{Yoo2012,Meng2016,Yu2018}, treat the control problem as a multi-player interactive game in order to find a solution that benefits all parties. Situation-aware planning typically utilizes a partially observable Markov decision process (POMDP) to account for the uncertainties in object vehicle (OV) actions and how the ego vehicles decisions impact them \cite{Liu2015}. MPC and situation-aware approaches are not, however, mutually exclusive. For example, in \cite{Zhou2018} a POMDP OV prediction algorithm is utilized in conjunction with a multi-policy MPC to account for interactions between agents. Although such approaches could potentially improve realism, they are hardly tractable for the high volume traffic evaluations we seek to do here. So, in this paper, we formulate a DMPC approach, along with other distributed state prediction and speed assignment functions, that can be applied to CAVs operating in mixed traffic, and then proceed with evaluating the traffic efficiency gains.

\vspace{-0.15in}
\subsection{Analysis of Benefits of CAVs in Traffic}\label{sec:Benefits}

From early works on automated highway systems, such as \cite{Rao1994}, it was shown that throughput or capacity could be significantly increased and delays from incidents blocking lanes could be mitigated with full penetration of CAVs. In general the underlying mechanism that realizes these gains is speed harmonization (SH). Currently implemented techniques for SH that rely on human drivers have produced mixed results \cite{Bham2010,Chang2011} due to the lack of compliance \cite{Talebpour2013}. The introduction of CAVs into the market presents a promising opportunity to realize the full potential of SH. In the literature, there tends to be two groups of SH techniques through individual control of CAVs: the first (indirect SH) indirectly harmonizes speed by reducing accelerations, and/or tracking a fixed headway (either distance or time) to the preceding vehicle \cite{WangM2016,Ard2020,RiosTorres2018}; the second (direct SH) attempts to directly harmonize speed by tracking either the current estimated average velocity of traffic \cite{Ngoduy2009,Ma2016,Goulet2020} or an optimal, in some sense, velocity based on the current traffic state, including velocity, density and/or demand \cite{WangM20162}. We will briefly review some of the specific implementations and their reported benefits of the two SH methods in the following paragraphs.

The indirect SH methods in \cite{WangM2016} and \cite{Ard2020} both balance minimizing accelerations and tracking a desired distance headway, while ensuring safety. Through simulations on a 3.65km single lane link, \cite{Ard2020} showed the ability of the presented MPC-based cooperative cruise control to increase throughput by $5\%$ at a CAV penetration as low as $30\%$. Fuel consumption at all CAV penetrations and traffic demands is also reduced, peaking at more than 25\% with a traffic demand of 2000 veh/hour and 100\% CAV penetration. The cooperative car-following controller in \cite{WangM2016} is coupled with a human model for lane change in order to consider a 2 lane highway with artificial bottlenecks created by reducing speed limits. The simulations therein show a 10\% to 15\% increase in throughput and a $3\%$ to $15\%$ decrease in fuel consumption when varying the CAV penetration from 5\% to 100\%. Although such results are promising, the resulting travel speed of a given CAV is dependent purely on the preceding vehicle's velocity and their underlying control scheme. This means there is potential for further benefits by more intelligently choosing the harmonization speed through direct SH.

The car-following model from \cite{WangM2016} is coupled with a decentralized variable speed limit algorithm for direct SH in \cite{WangM20162} by adding a desired velocity tracking term with the purpose of resolving traffic jams. The associated traffic flow and fuel consumption results in \cite{WangM20162} are, however, similar to that in \cite{WangM2016} as the desired velocity tracking term is only active during cruising, which is rarely the operating mode in dense traffic. Our prior work in \cite{Goulet2020} compared the 2D maneuver planning DMPC proposed here in two different SH configurations, an indirect SH (tracking immediate neighbors velocity) and direct SH (tracking estimated average traffic speed). The results therein showed the potential for direct SH to improve travel time by as much as 15\% and fuel consumption by around 6\% over indirect SH. Travel time reduction translates to an increase in throughput. Two different direct SH methods in \cite{Ngoduy2009} and \cite{Ma2016} assume that a given CAV receives the current downstream jam velocity from a roadside measurement unit (in \cite{Ma2016}) or preceding connected vehicle (in \cite{Ngoduy2009}) and then linearly reduces its speed to match.  Through continuum modeling, \cite{Ngoduy2009} shows a potential capacity increase of around 22\% at 50\% CAV penetration. The work in \cite{Ma2016} utilized control vehicles in real world traffic to implement the linear SH method. Measurements from preceding and trailing probe vehicles showed a reduction in speed variations/oscillations after the control vehicles, however, they did not directly correlate this to a travel time improvement. A drawback to such methods is the assumption that downstream velocity information is always available, which may not be the case. For this reason, in our work, we choose to estimate the current traffic speed on-board in a distributed fashion, where CAVs use measurements of neighboring vehicles from on-board sensors and, if available, improve the estimate with information shared from neighboring CAVs.

With the exception of our work in \cite{Goulet2020}, the previously discussed SH methods either do not account for lane changes or assume the lane changes are handled by human drivers, without taking advantage of the coupling between longitudinal and lateral decisions. Although there are multiple proposals in the literature for 2D control of CAVs on multi-lane roads, as reviewed in Section \ref{sec:MotionPlanning}, few investigate the traffic flow and fuel consumption benefits in large scale traffic simulations. In \cite{Dollar2018}, small scale simulations with four CAVs and one impeding vehicle show the potential for the MIP-based MPC to reduce fuel consumption by 8.4\% and travel time by 6.2\% over a baseline rule-based lane selection algorithm coupled with the intelligent driver model. Our prior conference papers \cite{Goulet2019} and \cite{Goulet2020} showed promise for a 2D maneuver planning DMPC to reduce travel time and fuel consumption as CAV penetration increases, however, the benefits associated with planning in 2D over 1D was not investigated. There are, however, some works that look at the benefits of selecting optimal lanes without coupled longitudinal planning \cite{Jin2014,Kang2018}. For example, in \cite{Kang2018} optimal lane selection with human longitudinal control realizes travel time reduction of 1\% at 20\% penetration and up to over 10\% at 100\% penetration on a 5 lane 4 km long highway with a single on-ramp.

\vspace{-0.05in}
\section{Details of the Proposed Coordinated and Distributed Control Framework}\label{sec:ControlFramework}

\begin{figure}[t]
\parbox{3.375in}{\centering\includegraphics[width=2.5in]{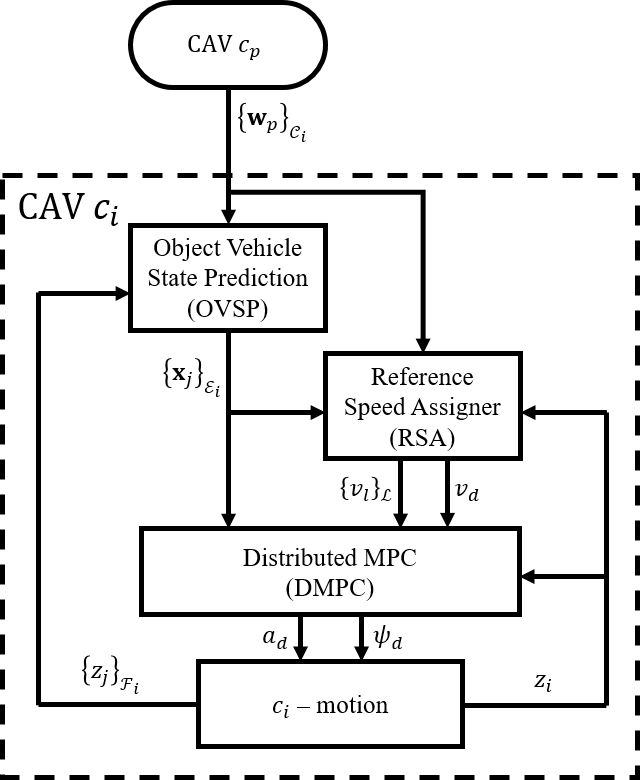}}
\caption{Control Framework for CAV $c_i$ where $\left\{\mathbf{w}_p\right\}_{\mathcal{C}_i}$ is the set of information matrices from each CAV $c_p$ communicating with $c_i$, $\mathbf{x}_i$ is the predicted optimal state trajectory for $c_i$ (not shown), $\left\{\mathbf{x}_j\right\}_{\mathcal{E}_i}$ defines the set of predicted state trajectories for OVs in the extended neighborhood of $c_i$, $v_d$ is the desired velocity of $c_i$, $\left\{v_l\right\}_{\mathcal{L}}$ is the set of reference velocities for each lane on the current link, $z_i$ are the measurements about $c_i$, $\left\{z_j\right\}_{\mathcal{F}_i}$ is the set of measurements about OVs in the FOV of $c_i$, and $a_d$ and $\psi_d$ are the control inputs, respectively, desired \vspace{-0.1in} acceleration and desired deviation in yaw angle.}
\label{fig:ctrl_framework} 
\end{figure}
Fig. \ref{fig:ctrl_framework} shows the overview of the proposed framework. For a given CAV $c_i$ there are three main components/blocks to the control framework that we propose and analyze in this paper; the object vehicle state prediction (OVSP) block, the reference speed assigner (RSA) block, and the DMPC block. Prior to outlining the flow of the proposed framework, we will define three sets of vehicles relevant to CAV $c_i$; $\mathcal{C}_i$ is the set of all CAVs $c_p \left(p \neq i\right)$ communicating with CAV $c_i$, $\mathcal{F}_i$ is the set of all OVs $ov_j$ in the field of view (FOV) of $c_i$, and $\mathcal{E}_i = \mathcal{C}_i \cup \mathcal{F}_i$ is the extended neighborhood of CAV $c_i$. It is possible that $\mathcal{C}_i\cap\mathcal{F}_i \neq \emptyset$. 
\begin{figure}[t]
\parbox{1in}{\centering\includegraphics[width=3.45in]{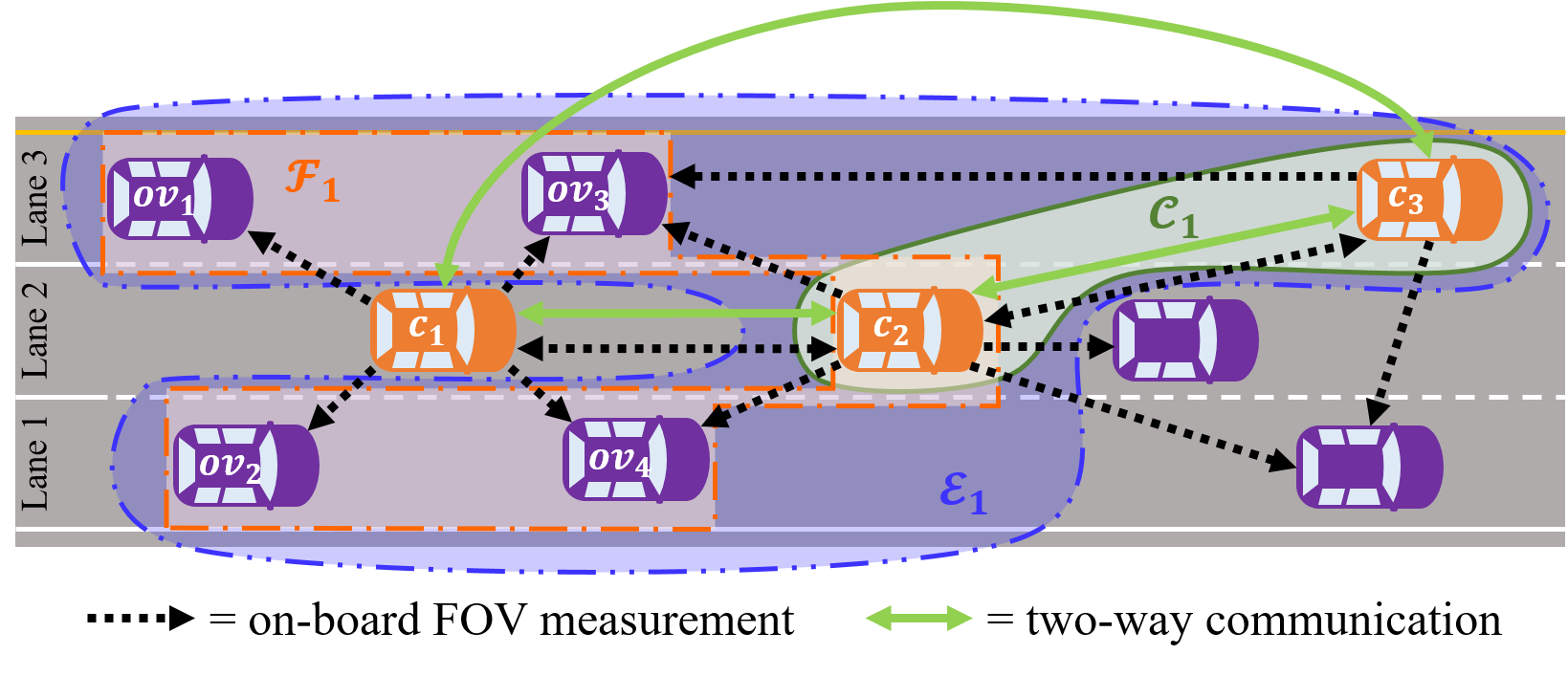}}
\caption{Example traffic topology outlining $\mathcal{C}_1= \left\{c_2, \ c_3\right\}$ the sets of vehicles communicating with $c_1$, $\mathcal{F}_1 = \left\{ov_1, \ c_2, \ ov_2, \ ov_3, \ ov_4\right\}$ the set of vehicles within the FOV of $c_1$, and $\mathcal{E}_1 = \left\{ov_1, \ ov_2, \ c_2, \ c_3, \ ov_3, \ ov_4\right\}$ the set of vehicles within the extended neighborhood of $c_1$. }
\label{fig:traffic_topology}
\end{figure}
Fig. \ref{fig:traffic_topology} depicts a potential traffic topology and clarifies the difference between $\mathcal{C}_i$, $\mathcal{F}_i$, and $\mathcal{E}_i$. The indices $i$, $j$, and $p$ are integers. Other symbols are defined in the caption of Fig. \ref{fig:ctrl_framework} and in the rest of this document.

The OVSP block takes as input the information matrix $\left\{\mathbf{w}_p\right\}_{\mathcal{C}_i}$ shared by each CAV $c_p\in\mathcal{C}_i$, and the measurements of all OVs in the set $\mathcal{F}_i$. Let $\mathbf{w}_p$ contain the motion plan of CAV $c_p$ and other local traffic information. The OVSP block then combines, associates, and reindexes the data for all vehicles in the set $\mathcal{E}_i$. The output of the OVSP block is the current (estimated) and predicted future state trajectory of each OV in the set $\mathcal{E}_i$ for use by the DMPC and RSA blocks. If a given OV, $ov_j$, is in both $\mathcal{C}_i$ and $\mathcal{F}_i$, a practical data association method may be adopted, as in the collaborative perception framework outlined in \cite{Yoon2019}. However, here we treat OVSP for CAVs and HDVs separately as will be detailed in Section \ref{sec:OV_pred}.

The RSA block takes in the state estimates of each $ov_j$, communicated information, and ego-measurements at CAV $c_i$ in order to estimate the instantaneous average speed of surrounding traffic. Using this estimate and based on a protocol to be described in Section \ref{sec:RSA}, the RSA block assigns the desired speed $v_d$ for CAV $c_i$ and a reference speed $v_{l}$ to each lane $l \in \mathcal{L}$, where $\mathcal{L}$ is the set of lanes on the current link. To clarify the difference between $v_d$ and $v_l$: the desired speed $v_d$ is the speed on a given link required to meet the desired travel time of CAV $c_i$, while the lane reference speed $v_l$ is the speed supported by the given lane $l$. 

The DMPC block takes in ego measurements $z_i$ and the outputs from the OVSP and RSA blocks to find the optimal 2D maneuver plan over a finite horizon. The formulation of the DMPC will be discussed in detail in the following subsection. The first step control inputs (desired tangential acceleration $a_d$ and desired deviation in yaw angle $\psi_d$) are then passed to lower level vehicle dynamics controllers.

\vspace{-0.15in}
\subsection{Distributed Model Predictive Control Formulation}\label{sec:MPC}

The DMPC problem to be solved at each CAV $c_i$ can be posed compactly as follows:
\begin{subequations}\label{eq:MPC}
\begin{gather}
\min_{\mathbf{u}_i}\left[\left\|\mathbf{F}_i\right\|^2_{\mathbf{P}_{f,i}}+\left\|\mathbf{G}_i\right\|^2_{\mathbf{P}_{g,i}}+\left\|\mathbf{H}_i\right\|^2_{\mathbf{P}_{h,i}}+\left\|\mathbf{u}_i\right\|^2_{\mathbf{R}_i}\right] \label{eq:cost}\\
s.t. \ \ x_{i,k+1} = f\left(x_{i,k},u_{i,k}\right), \ x_{i,k} \in X_i, \ u_{i,k} \in U_i \label{eq:model}\\
x_{i,0} = \hat{x}_{i,0}\label{eq:initial_condition}\\
c(x_{i,k},u_{i,k}) \geq 0 \label{eq:const}\\
\begin{matrix}g(x_{i,k},x_{j,k}) \geq 1 & \forall \ ov_j \in \mathcal{E}_i\end{matrix}\label{eq:collision_const},
\end{gather}
\end{subequations}
where 
the state vector $x_{i,k}$, control vector $u_{i,k}$ and nonlinear motion dynamics model $f\left(x_{i,k},u_{i,k}\right)$ in \eqref{eq:model} are to be defined below. The definition of the admissible sets for the states, $X_i$, and inputs, $U_i$, along with other road friction and boundary constraints in the function $c(x_{i,k},u_{i,k})$ of \eqref{eq:const}, and obstacle avoidance constraints in the function $g(x_{i,k},x_{j,k})$ of \eqref{eq:collision_const}, will be detailed in Section \ref{sec:constraints}. The problem is further subject to the current state estimate $\hat{x}_{i,0}$ in \eqref{eq:initial_condition}. The cost function is divided into four components that will be discussed shortly below: the lane-dependent cost $\|\mathbf{F}_i\|^2_{\mathbf{P}_{f,i}}$, lane-independent cost $\|\mathbf{G}_i\|^2_{\mathbf{P}_{g,i}}$, predictability cost $\|\mathbf{H}_i\|^2_{\mathbf{P}_{h,i}}$ and input cost $\|\mathbf{u}_i\|^2_{\mathbf{R}_i}$. Each cost term is defined as the weighted 2-norm of the vector (outputs $\mathbf{F}_i$, $\mathbf{G}_i$, $\mathbf{H}_i$ or input $\mathbf{u}_i$), with a respective symmetric and positive semi-definite weighting matrix ($\mathbf{P}_{f,i}$, $\mathbf{P}_{g,i}$, $\mathbf{P}_{h,i}$, or $\mathbf{R}_i$). Notation wise, $\|\mathbf{F}\|^2_\mathbf{P} = \mathbf{F}^T\mathbf{P}\mathbf{F}$. Going forward, where there is no ambiguity, we will omit the time index $k$ and the ego-vehicle index $i$ in order to reduce clutter.

\subsubsection{Dynamics Model}\label{sec:ego_model}

\begin{figure}[t]
\parbox{3.375in}{\centering\includegraphics[width=3.2in]{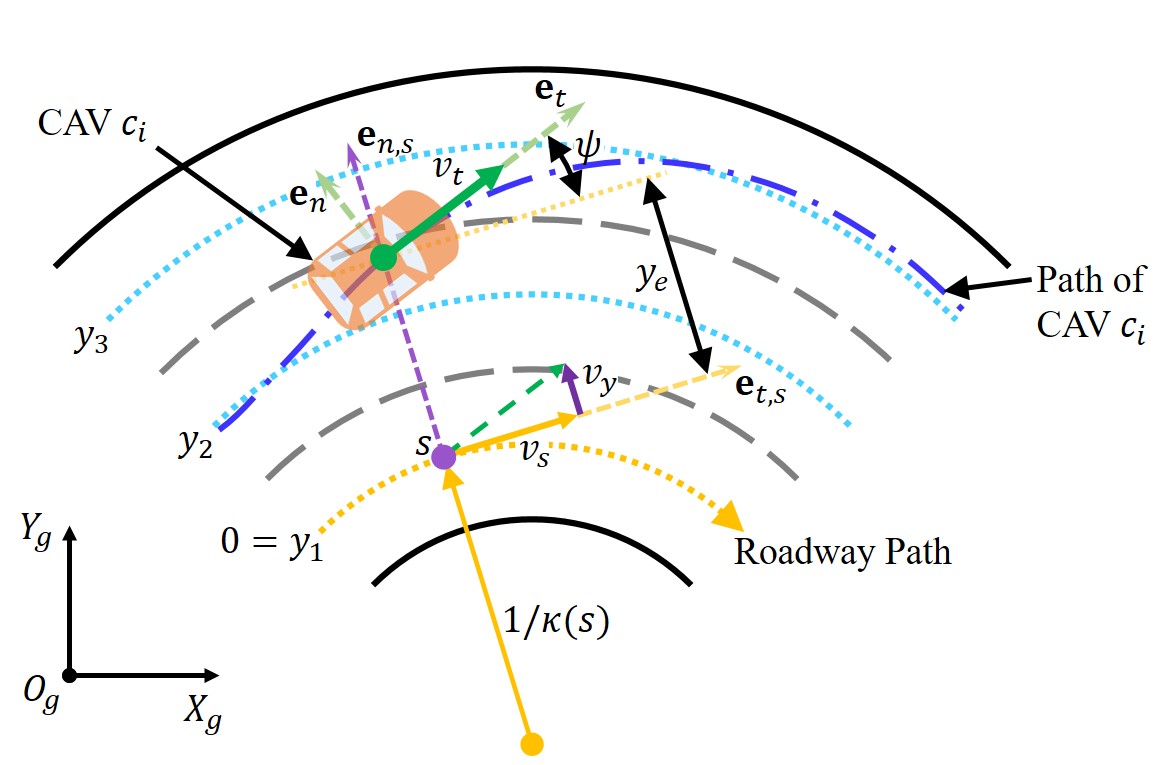}}
\caption{Ego vehicle motion in the Frenet frame where $O_g$ is the global origin, $X_g$ and $Y_g$ are the axis of the global coordinate frame, $\mathbf{e}_n$ and $\mathbf{e}_t$ are the coordinate axes of the normal-tangential (n-t) coordinate system aligned with CAV $c_i$'s path, whereas $\mathbf{e}_{n,s}$ and $\mathbf{e}_{t,s}$ are the coordinate axes of the n-t coordinate system aligned with the roadway path. Further, $v_s$ and $v_y$ are the components of $v_t$ in the $\mathbf{e}_{t,s}$- and $\mathbf{e}_{n,s}$-directions, respectively, and $\kappa\left(s\right)$ is the roadway curvature as a function of the roadway path coordinate $s$.}
\label{fig:ego_motion} 
\end{figure}

The ego-vehicle is assumed to follow a non-linear particle motion model expressed in the path intrinsic Frenet frame. However, it is possible to adopt other models for this vehicle level dynamics \cite{Weiskircher2017}. The state vector is $x = \begin{bmatrix}s&y_e&v_t&\psi&a_t\end{bmatrix}^T$, where $s$ is the position along the roadway path, $y_e$ is the lateral deviation from the roadway path, $v_t$ is the tangential velocity, $\psi$ is the deviation in heading from the roadway orientation, and $a_t$ is the tangential acceleration. Fig. \ref{fig:ego_motion} depicts these states for clarity.
The motion dynamics model is then \cite{Weiskircher2017}:
\begin{equation}\label{eq:ego_model}
\dot{x} = \begin{bmatrix}\dot{s}\\
\dot{y}_e\\
\dot{v}_t\\
\dot{\psi}\\
\dot{a}_t\end{bmatrix} = \begin{bmatrix}\frac{v_t}{1-y_e\kappa\left(s\right)}\cos{\psi}\\
v_t\sin{\psi}\\
a_t\\
\tau_{\psi}\left(\psi_d-\psi\right)\\
\tau_{a}\left(a_d-a_t\right)\\\end{bmatrix}.
\end{equation}
The vehicle's acceleration and deviation in heading are assumed to follow the desired inputs, $a_d$ and $\psi_d$, with first order dynamics and time constants $\tau_a$ and $\tau_{\psi}$, respectively.  The above formulation in \eqref{eq:ego_model} is discretized via a $4^{th}$-order Runge-Kutta explicit integrator in the solver for the DMPC problem \eqref{eq:MPC}. There are additional states required for computational modeling that will be introduced and discussed in the remainder of Section \ref{sec:ControlFramework}. For the complete augmented state dynamics, please see Appendix \ref{sec:aug_state}.

\subsubsection{Cost}\label{sec:cost}

The lane-dependent cost is conceived to penalize both the lane centerline and lane reference speed  tracking error over the prediction horizon. Specifically, the lane dependent cost for the ego vehicle is the weighted 2-norm of the vector $\mathbf{F} = \begin{bmatrix}F_1&\cdots&F_{N_h}\end{bmatrix}^T$ with weighting matrix $\mathbf{P}_f = \diag\begin{Bmatrix}P_{f,1},&\cdots,&P_{f,N_h}\end{Bmatrix}$, where $N_h$ is the number of steps in the prediction horizon. The lane-dependent output vector F at each time step is defined as:
\vspace{-0.05in}
\begin{multline}\label{eq:lane_dep}
F = \begin{matrix}[d_{1}\left(y_{e}-y_{1}\right) & \cdots & d_{N_l}\left(y_{e}-y_{N_l}\right) \end{matrix}\\
\begin{matrix}d_{1}\left(v_t-v_{1}\right) & \cdots & d_{N_l}\left(v_t-v_{N_l}\right)]\end{matrix}^T
\end{multline}
where $d_l$ is the lane decision variable of lane $l$, $y_l$ is the position of the centerline of lane $l$, and $N_l$ is the total number of lanes in the set $\mathcal{L}$. The lane decision variables are constrained as follows:
\begin{subequations}\label{eq:dl_constraints}
\begin{gather}
\begin{matrix}d_l \in \begin{bmatrix}0,&1\end{bmatrix}&\forall l \in \begin{Bmatrix}1,&...,&N_l\end{Bmatrix}\end{matrix}\label{eq:dl_set}\\
\sum_{l=1}^{N_l}d_l = 1\label{eq:dl_initial_const}\\ \intertext{from which,}
\vspace{-0.05in}
d_{N_l} = 1 - \sum_{l=1}^{N_l-1}d_l,\label{eq:dNl_const}.
\end{gather}
\end{subequations}
We impose the integrator dynamics on the remaining lane variables \cite{Wang2019}:
\begin{equation}\label{eq:dl_dynamics}
\begin{matrix}\dot{d}_l = u_{d_l} &\forall \ l \in \left\{1,...,N_l-1\right\}\end{matrix},
\end{equation}
where $u_{d_l}$ is the control input for a given lane decision variable $d_l$. The combination of constraints in \eqref{eq:dl_constraints} guarantees that the lane decision variables are normalized while reducing the number of manipulated variables $u_{d_l}$ by 1.

In a similar manner, the lane-independent cost is defined as the weighted 2-norm of the vector $\mathbf{G} = \begin{bmatrix}G_1&\cdots&G_{N_h}\end{bmatrix}^T$ with weighting matrix $\mathbf{P}_g = \diag\begin{Bmatrix}P_{g,1},&\cdots,&P_{g,N_h}\end{Bmatrix}$. The lane-independent output vector $G$ consists of a desired velocity tracking term, a slack state tracking term, and a term to avoid choosing multiple lanes at each time step in the horizon:
\begin{equation}\label{lane_ind}
G = \begin{bmatrix}v_t-v_d & \zeta-v_d & 1-\sum_{l=1}^{N_l}d_l^2\end{bmatrix}^T,
\end{equation}
where $\zeta$ is a slack state variable for ensuring a comfortable following distance during normal driving. The function of $\zeta$ will be discussed further in Section \ref{sec:constraints}. 

As CAV $c_i$ will be sharing its calculated motion plan, so that each communicating CAV $c_p \in \mathcal{C}_i$ can compute the drivable space over their prediction horizon, it is necessary that CAV $c_i$ acts in a predictable manner. To this end, a predictability cost has been added that penalizes deviations from the prior plan with respect to $s$ and $y_e$. The predictability cost is the weighted 2-norm of the vector $\mathbf{H} = \begin{bmatrix}H_1&\cdots&H_{N_h-1}\end{bmatrix}^T$ with weighting matrix $\mathbf{P}_h = \diag\begin{Bmatrix}P_{h,1},&\cdots,&P_{h,N_h-1}\end{Bmatrix}$, where the output vector $H$ at each time step is:
\begin{equation}\label{eq:predictability_cost}
H = \begin{bmatrix}s-s^- & y_e - y_e^-\end{bmatrix}^T.
\end{equation}
The superscript $^-$ denotes a synchronized prior plan. The plans are synchronized via a constant velocity model assuming CAV $c_i$ has traveled according to the plan it shared from the time it was made until the current time. The approach used for synchronization is described in Section \ref{sec:cav_pred} below.

Lastly, we define the input cost as the 2-norm of the augmented input vector $\mathbf{u} = \left[\begin{matrix}u_1&\cdots&u_{N_h-1}\end{matrix}\right]^T$ with the weighting matrix $\mathbf{R} =\diag\left\{\begin{matrix}R_1&\cdots&R_{N_h-1}\end{matrix}\right\}$. The input vector at each time step is $u = \begin{bmatrix}a_{d}&\mkern-3mu\delta\psi_d&\mkern-3mu u_\zeta&\mkern-3mu u_{d_1}&\mkern-3mu\cdots&\mkern-3mu u_{d_{N_l-1}}\end{bmatrix}^T$. The choice to penalize all inputs is made in order to obtain practically smooth solutions. For example, in order to penalize the rate of change of the heading deviation $\delta\psi_d$, the integrator dynamics $\dot{\psi_d} = \delta\psi_d$ are introduced. We note that it is unnecessary to penalize $\psi_d$ directly, as in order to minimize the lane centerline tracking term in \eqref{eq:lane_dep}, the vehicle will tend to follow the heading of the lane. The manipulated input $u_{\zeta}$ controls the slack state $\zeta$, where $\dot{\zeta} = u_{\zeta}$.

\subsubsection{Constraints}\label{sec:constraints}

Constraints relevant to the limits of the vehicle include the friction ellipse \eqref{eq:friction_ellipse} and the minimum turning radius \eqref{eq:turning_rad}:
\begin{subequations}\label{eq:vehicle_const}
\begin{gather}
\left(\frac{a_n}{\eta}\right)^2 + a_t^2 \leq \left(\mu g\right)\label{eq:friction_ellipse}\\
v_t \kappa\left(s\right) + \dot{\psi} \leq v_t \upperb{\kappa}\label{eq:turning_rad}
\end{gather}
\end{subequations}
where $a_n = v_t^2 \kappa\left(s\right) + \dot{\psi}v_t$ is the normal acceleration, $\eta$ is a normalization factor, $\mu$ is the coefficient of friction, $g$ is acceleration due to gravity, $\dot{\psi}$ is the yaw rate, which may be modeled with \eqref{eq:ego_model},
and $\upperb{\kappa}$ is the permissible curvature based on the minimum turning radius of the vehicle, as defined in \cite{Weiskircher2017}.

We also impose the following box-type constraints:
\begin{subequations}
\begin{gather}
s \leq \upperb{s}\\
\lowerb{y}_e \leq y_e \leq \upperb{y}_e\\
 0 \leq v_t \leq \upperb{v}_t,
\end{gather}
\end{subequations}
where $\upperb{s}$ is the upper limit on the progress along the path of CAV $c_i$, which may be assigned at a higher level for imposing traffic signals or stop signs \cite{Weiskircher2017}, $\lowerb{y}_e$ and $\upperb{y}_e$ are the right and left roadway bounds, respectively, and $\upperb{v}_t$ is the speed limit.

To ensure that the computed plan guarantees CAV $c_i$  can remain within the roadway bounds beyond the prediction horizon, an additional terminal constraint is added (see Fig. \ref{fig:ye_term_const} for construction):
\begin{figure}[t]
\parbox{3.375in}{\centering\includegraphics[width=2.4in]{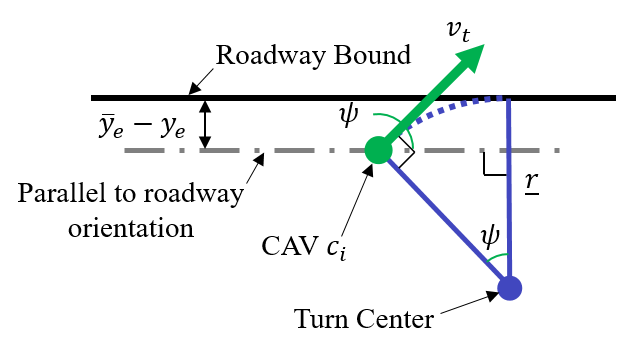}}
\caption{Schematic depicting derivation of the terminal constraint on the upper bound of $y_e$ (an analogous derivation is used for the lower bound).}
\label{fig:ye_term_const} 
\end{figure}
\begin{equation}
\lowerb{y}_e \mkern-6mu + \lowerb{r}_{N_h} \mkern-6mu \left(1 \mkern-3mu - \mkern-3mu \left|\cos{\psi_{N_h}} \mkern-2mu \right|\right) \mkern-2mu \leq y_{e,{N_h}} \mkern-4mu \leq \upperb{y}_e \mkern-6mu - \lowerb{r}_{N_h} \mkern-6mu \left(1 \mkern-6mu - \mkern-3mu \left|\cos{\psi_{N_h}} \mkern-2mu \right|\right),
\end{equation}
where $\lowerb{r}_{N_h} = v_{t,N_h}^2/\upperb{a}_n$ is the minimum permissible turning radius based on the tangential velocity $v_{t,N_h}$ at the end of the prediction horizon and the normal acceleration limit $\upperb{a}_n$.

Next, we detail the collision avoidance constraints between CAV $c_i$ and $ov_j$ (given compactly in 1e). These are formulated as hyperelliptical constraints centered at the location of $ov_j$ and based on the size and relative orientation of CAV $c_i$ and $ov_j$ (see Appendix \ref{sec:ov_ellipse} for the construction): 
\begin{equation}\label{eq:ellipse}
\left(\frac{y_{e,i} - y_{e,j}}{\gamma_{ij}}\right)^4 + \left(\frac{s_i-s_j}{\lambda_{ij}+\lambda_{b,ij}+\beta\zeta}\right)^4\geq1.
\end{equation}
In \eqref{eq:ellipse}, $\gamma_{ij}$ and $\lambda_{ij}$ are the half minor (lateral) and half major (longitudinal) axes, respectively, of the hyperelliptical constraint between CAV $c_i$ and $ov_j$. The term $\beta\zeta$ is the distance slack with fixed comfort time headway $\beta$ and slack state $\zeta$ (with units of velocity). We will note that $\zeta$ is constrained in $\begin{bmatrix}0,&\upperb{v}_t\end{bmatrix}$ to ensure the following distance remains positive and does not become too large. The term $\lambda_{b,ij}$ ensures that if the lead vehicle, whether that is CAV $c_i$ or $ov_j$, enters an emergency braking situation, there is a sufficient distance to the trailing vehicle for both to stop without collision. Assuming both CAV $c_i$ and $ov_j$ follow a constant acceleration model, and their maximum allowable decelerations $\lowerb{a}_{s}$ in the $\mathbf{e}_{t,s}$-direction (Fig. \ref{fig:ego_motion}), are known, the minimum distance between CAV $c_i$ and $ov_j$ required for both vehicles to come to a stop without collision is:
\begin{equation}\label{eq:brake_safe}
\lambda_{b,ij} = \begin{cases}
\frac{1}{2}\left(\frac{v_{s,i}^2}{\lowerb{a}_{s,i}}-\frac{v_{s,j}^2}{\lowerb{a}_{s,j}}\right)&\text{if}\left(s_i-s_j\right)\left(v_{s,i}-v_{s,j}\right)<0\\
0&\text{otherwise},
\end{cases}
\end{equation}
where $a_s$ is the component of $a_t$ in the $\mathbf{e}_{t,s}$-direction. The conditional statement in \eqref{eq:brake_safe} guarantees that $\lambda_{b,ij}$ is considered only if CAV $c_i$ is ahead of and traveling slower than $ov_j$, or behind and traveling faster.

In the lateral direction, to prevent a collision from happening beyond the end of the horizon, $\gamma_{ij,N_h}$ is computed as follows:
\begin{subequations}\label{eq:lat_term_const}
\begin{gather}
\gamma_{ij,N_h} = \gamma_{ij} + \begin{cases}
\gamma_{c,ij} & \text{if } \sign{v_{y,j}} = \sign{\left(y_{e,i} - y_{e,j}\right)} \\
0& \text{otherwise}
\end{cases}\\
\gamma_{c,ij} = \begin{cases}
\frac{\left(v_{y,i} - v_{y,j}\right)^2}{2\left(\upperb{a}_{y,i}-a_{y,j}\right)}&\text{if } t_{c,ij} = \frac{v_{y,i}-v_{y.j}}{\upperb{a}_{y,i}-a_{y,j}} < 0\\
M&\text{otherwise}
\end{cases}
\end{gather}
\end{subequations}
where $a_{y}$ is the component of $a_t$ and $\upperb{a}_{y}$ is the limit of acceleration, both in the $\mathbf{e}_{n,s}$-direction. The term $\gamma_{c,ij}$ ensures CAV $c_i$ has sufficient space to reach a zero relative velocity with respect to $ov_j$, with $t_{c,ij}$ being the time required to reach zero relative velocity in the lateral direction, and $M$ being a sufficiently large number to ensure CAV $c_i$ is not alongside $ov_j$ if no feasible $\gamma_{c,ij}$ exists.

\subsection{Object Vehicle State Prediction}\label{sec:OV_pred}

The OVSP block treats CAVs and HDVs differently. The CAVs are assumed to share a previously calculated plan, whereas the states of HDVs are estimated via decoupled longitudinal and lateral dynamics models. We discuss the two treatments separately below.

\subsubsection{Connected and Automated Vehicles}\label{sec:cav_pred}

The CAV $c_p$ communicating with CAV $c_i$ sends an information matrix $\mathbf{w}_p = \begin{Bmatrix}\mathbf{x}_p&t_p&\left\{\begin{bmatrix}\mu_{v,lp}&\lowerb{\alpha}_{lp}&\upperb{\alpha}_{lp}&N_{v,lp}\end{bmatrix}\right\}_{l\in\left\{1,...,N_l\right\}}\end{Bmatrix}$, where the shared state prediction $\mathbf{x}_p = \left\{\begin{bmatrix}s_{p,k_p} & y_{e,p,k_p}\end{bmatrix}_{k_p=0:N_{h,p}}^T\right\}$ consists of the predicted position along the path $s_p$, and lateral error $y_{e,p}$ of CAV $c_p$. Time index $k_p=0$ corresponds to $t_p$, the discrete time at which the plan was made by CAV $c_p$. The remaining elements in $\mathbf{w}_p$ are defined with, and to be utilized by, the RSA in Section \ref{sec:RSA}.

As the controller update step $\delta t_s$ and the time step $\delta t_h$ within the prediction horizon may not be equal, it is necessary to synchronize the plan shared by CAV $c_p$ with the planning cycle in CAV $c_i$. 
\begin{figure}[t]
\parbox{3.5in}{\centering\includegraphics[width=3.5in]{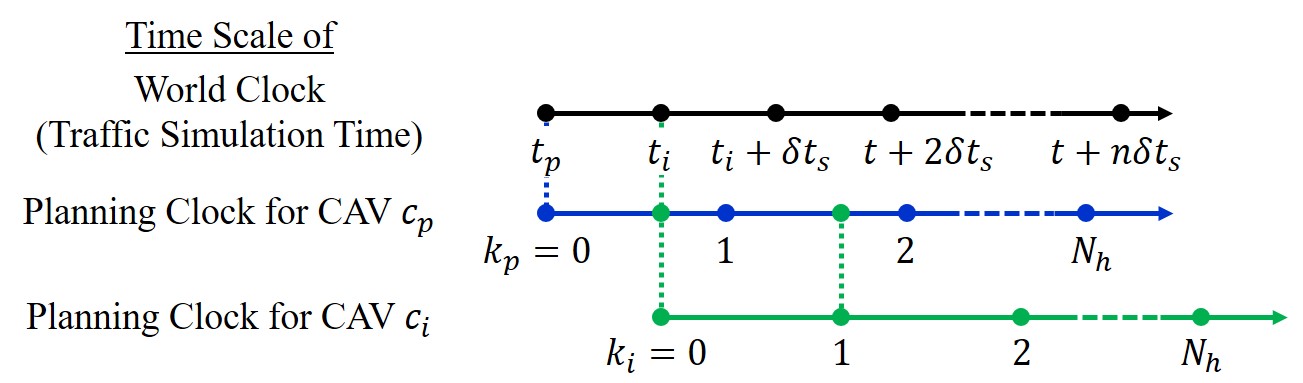}}
\caption{Time scale illustrating the need to synchronize the plan shared by $c_p$.}
\label{fig:synchronization} 
\end{figure}
Fig. \ref{fig:synchronization} illustrates the difference in time scales, where $k_i$ and $k_p$ are the discrete time indices for the prediction horizon of CAV $c_i$ and CAV $c_p$, respectively. Further, $k_i=0$ corresponds to the discrete time $t_i$ when CAV $c_i$ is planning.

We utilize the following assumptions to synchronize the shared plans:
\begin{itemize}
\item The communicating CAV $c_p$ is assumed to follow the shared plan from $t_p$ until $t_i$.
\item Both CAV $c_i$ and CAV $c_p$ utilize the same MPC discretization, i.e. $N_h \mkern-3mu = \mkern-3mu N_{h,i} \mkern-3mu = \mkern-3mu N_{h,p}$ and $\delta t_h \mkern-3mu = \mkern-3mu \delta t_{h,i} \mkern-3mu = \mkern-3mu \delta t_{h,p}$.
\item The states $s_p$ and $y_{e,p}$ are decoupled and follow constant velocity kinematic models between time steps.
\item The delay $t_i-t_p \leq \delta t_h$.
\item The controller update step $\delta t_s \leq \delta t_h$.
\end{itemize}
The synchronization for $s_p$ is then completed using the following equations:
\begin{subequations}\label{eq:lin_interp}
\begin{gather}
\begin{matrix}s_{p,k_i} = \hat{s}_{p,k_i}& \text{for } k_i=0\end{matrix}\\
\Delta s_{p,k_p} = \begin{cases}s_{p,k_p}-s_{p,k_p-1} & \text{if } k_p = N_h\\
s_{p,k_p+1}-s_{p,k_p}& \text{otherwise}\end{cases}\\
s'_{p,k_p} = s_{p,k_p} + \frac{t_i-t_p}{\delta t_h}\Delta s_{p,k_p}\\
\begin{matrix}s_{p,k_i}^- = s'_{p,k_p} + s_{p,k_i=0} - s'_{p,k_p=0}\\
\text{for } k_i=k_p \in \left\{1,2,...,N_h\right\}\end{matrix},
\end{gather}
\end{subequations}
where $\hat{s}_{p,k_i}$ is the latest estimate of CAV $c_p$'s position at $k_i=0$, or time $t_i$, based on shared information and, if available, measurements from sensors on-board CAV $c_i$. Let $s'_{p,k_p}$ be the time shifted position of CAV $c_p$ along the path, while the term $s_{p,k_i=0}-s'_{p,k_p=0}$ translates the synchronized plan so its start coincides with $\hat{s}_{p,k_i}$. The synchronized position along the path of CAV $c_p$ at time step $k_i$ is then $s^-_{p,k_i}$. We will note that it is possible to expand this formulation such that $\delta t_s$ may be greater then $\delta t_h$ via simple logic, however, additional considerations for CAV $c_p$'s behavior beyond its planning horizon will be required. The synchronization for $y_{e,p}$ is pursued similarly.

\subsubsection{Unconnected Vehicles}\label{sec:HD_pred}

Predicting the future states and intentions of HDVs based on their surrounding environment (roadway structure, interactions with other vehicles, etc.) is a complex and difficult research question in and of itself with many proposed solutions \cite{Lefevre2014}. It is, however, not our focus. Here, we will assume HDVs follow decoupled linear feedback models for their motion in the $n-t$ coordinate system aligned with the roadway path. In order to extend these models to predict over the horizon, we will utilize Kalman filtering and the idea of the most likely measurement to update the covariance without shifting the mean \cite{DuToit2012}. Other prediction methods may be suitable for implementation within this framework, for example, one could assume HDVs follow a POMDP as in \cite{Zhou2018}.

The HDV $ov_j \in \mathcal{F}_i\setminus\mathcal{C}_i$ are assumed to follow a simple linear feedback car-following model in the longitudinal direction based on tracking a safe distance, $d_s$, from the position $s_{f,j}$ of their respective lead vehicle, and the lead vehicle's velocity $v_{f,j}$:
\begin{subequations}\label{ov_long}
\begin{gather}
\begin{split}
x_{s,j,k+1} = 
&\begin{bmatrix}1&\delta t\\
-K_s&1-K_{v_s}\end{bmatrix}\begin{bmatrix}s_{j,k}\\
v_{s,j,k}\end{bmatrix}\\
& + \begin{bmatrix}0&0&0\\
K_s &K_{v_s}&-K_s\end{bmatrix}\begin{bmatrix}s_{f,j,k}\\
v_{f,j,k}\\
d_s\end{bmatrix}
\end{split}\\
y_{s,j,k} = \begin{bmatrix}1&0\\
0&1\end{bmatrix}\begin{bmatrix}s_{j,k}\\
v_{s,j,k}\end{bmatrix},
\end{gather}
\end{subequations}
where the longitudinal state $x_{s,j}$ of $ov_j$ consists of its position  $s_j$ and velocity $v_{s,j}$ along its path. Let $K_s$ and $K_{v_s}$ be the distance tracking and velocity tracking gains, respectively, and $\delta t$ the time step. The parameter $d_s$ takes an assumed constant value, while $s_{f,j}$ and $v_{f,j}$ are assumed to be unknown disturbances and augmented into the state matrix via a constant velocity model for estimation purposes.

Similarly, it is assumed that $ov_j$ has a linear feedback controller to track its lane centerline $y_{ref,j}$ and a zero lateral velocity, as we presented in \cite{Wang2019}:
\begin{subequations}\label{ov_lat}
\begin{gather}
x_{y,j,k+1} = \begin{bmatrix}1&\delta t\\
-K_y&1-K_{v_y}\end{bmatrix}\begin{bmatrix}y_{e,j,k}\\
v_{y,j,k}\end{bmatrix}+\begin{bmatrix}0\\
K_y\end{bmatrix}y_{ref,j,k}\\
y_{y,j,k} = \begin{bmatrix}1&0\end{bmatrix}\begin{bmatrix}y_{e,j,k}\\
v_{y,j,k}\end{bmatrix}.
\end{gather}
\end{subequations}
The lateral state $x_{y,j}$ consists of the lateral position $y_{e,j}$ of $ov_j$, and the lateral velocity $v_{y,j}$ of $ov_j$. Let $K_y$ and $K_{v_y}$ be the lateral position and velocity tracking gains, respectively. Since it is not known which lane is being tracked by $ov_j$ at any given instant, the input $y_{ref,j}$ is assumed to be an unknown disturbance and augmented into the state matrix to be estimated via stationary dynamics.

\subsection{Reference Speed Assignment}\label{sec:RSA}

The lane reference speed will be assigned in a distributed manner by estimating the average speed of traffic in each lane via on-board measurements and shared information. First we will define CAV $c_i$'s estimated average velocity of traffic in lane $l$ as $\mu_{v,li}=\sum^{N_{v,li}}_{j=1}v_{t,j}/N_{v,li}$, where $N_{v,li}$ is the number of vehicles in lane $l$ in the FOV of CAV $c_i$. Then CAV $c_i$ will combine its estimate $\mu_{v,li}$ with each $\mu_{v,lp}$ estimate shared by communicating CAVs $c_p \in \mathcal{C}_i$. As mentioned in Section \ref{sec:cav_pred}, we assume the information matrix $\mathbf{w}_p$ shared by each CAV $c_p$ will contain a limited amount of information about the traffic environment.  The relevant shared information includes $\mu_{v,lp}$ and $N_{v,lp}$ for each lane $l$ at time $t_p$, and the upstream and downstream bounds of the FOV in each lane $l$, $\lowerb{\alpha}_{lp}$ and $\upperb{\alpha}_{lp}$, respectively.

To mitigate double counting vehicles, it is necessary to estimate the number of unique vehicles $N_{u,lp}$ in the FOV of CAV $c_p$. The following two assumptions allow us to estimate $N_{u,lp}$ using only the shared information:
\begin{itemize}
\item Each CAV's FOV in lane $l$ is a rectangle of equivalent width as the lane $l$ and length bounded by $\lowerb{\alpha}_{lp}$ and $\upperb{\alpha}_{lp}$,
\item OVs are evenly distributed within the FOV in lane $l$.
\end{itemize}
It should be noted that more elaborate descriptions of the FOV are possible at the cost of more complex computations\cite{Liu2018}.

CAV $c_i$ then estimates $N_{u,lp}$ as follows:
\begin{equation}
N_{u,lp} = \frac{A_{u,lp}}{A_{lp}} N_{v,lp},
\end{equation}
where $A_{u,lp}$ and $A_{lp}$ are, respectively, the unique and total FOV area in lane $l$ of CAV $c_p$. $A_{u,lp}$ is calculated based on a pairwise comparison of the FOV bounds $\lowerb{\alpha}_{lp}$ and $\upperb{\alpha}_{lp}$. The algorithm used is given in Appendix \ref{app:unique_FOV}. The lane reference velocity $v_l$ for the ego CAV $c_i$ is then estimated as:
\begin{equation}\label{eq:lane_ref_speed}
v_l = \frac{1}{N_{v,li} + \sum^{N_p}_{p=1}N_{u,lp}}\left[N_{v,li}\mu_{v,li} + \sum^{N_p}_{p=1}N_{u,lp}\mu_{v,lp}\right],
\end{equation}
 where $N_p$ is the number of CAVs communicating with CAV $c_i$.

The desired speed $v_d$ of CAV $c_i$ is then assigned based on the lane reference speed that would realize the closest travel time to the desired for the given link:
\begin{equation}\label{eq:vd}
v_d = \argmin_{v_l}{\left|v_l - L/T_d\right|}.
\end{equation}
Where $L$ is the length of the link and $T_d$ is the desired travel time on the link. One can solve this optimization problem using a simple search over the $N_l$ values.

When the local density of traffic $\rho_{l}$ in a given lane $l$  is low, it is undesirable to set the lane reference speed based on the average velocity. As an example, say there is one OV, $ov_j$, in lane $l$ traveling slower than the desired velocity of CAV $c_i$ yet behind CAV $c_i$. Lane $l$ would unnecessarily be penalized if $v_l$ is assigned by \eqref{eq:lane_ref_speed}, or to be the velocity of $ov_j$, while there is no downstream vehicle to impede CAV $c_i$'s motion on lane $l$. For this reason, if the density $\rho_{li}$ estimated by CAV $c_i$ is less than a predetermined threshold $\lowerb{\rho}$, the lane reference speeds are assigned using the rule-based speed assigner presented in \cite{Wang2019} and \cite{Goulet2019}, where:
\begin{equation}\label{eq:density_est}
\rho_{li} =  \frac{N_{v,li} + \sum^{N_p}_{p=1}N_{u,lp}}{\max{\left(\upperb{\alpha}_{li},\left\{\upperb{\alpha}_{lp}\right\}_{\mathcal{C}_i}\right)} - \min{\left(\lowerb{\alpha}_{li},\left\{\lowerb{\alpha}_{lp}\right\}_{\mathcal{C}_i}\right)}}.
\end{equation}

\section{Results and Discussions}\label{sec:Results}

\subsection{Evaluation Setup}\label{sec:EvalSetup}
In order to evaluate the impact of the proposed distributed maneuver planning and control framework, we implemented the framework in a traffic micro-simulation environment, by interfacing our custom code with the software VISSIM \cite{VISSIM}. The optimal control problem (OCP) of the DMPC for each CAV was solved via autogenerated code using the ACADO toolkit \cite{Houska2011}.

\begin{figure}[t]
\parbox{3.5in}{\centering\includegraphics[width=3.55in]{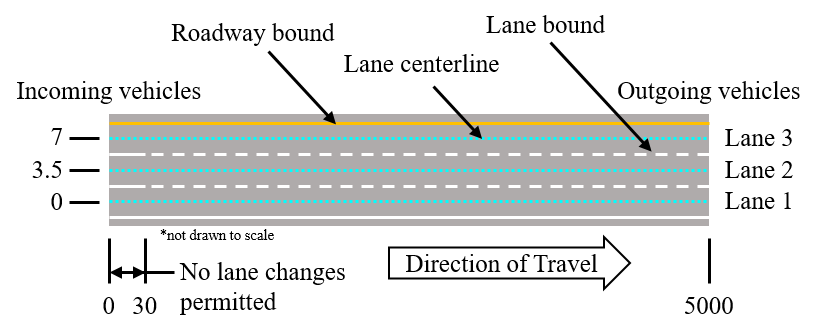}}
\caption{Schematic of the simulated  traffic network.}
\label{fig:network} 
\end{figure}

The simulated traffic network was comprised of a 5000m long straight three lane link, with a single input node and single output node as shown in Fig. \ref{fig:network}. 
The lanes are labeled from 1 to 3 with the rightmost lane being 1 and increasing to the left. Each lane is 3.5m wide with the origin of the $y_e$ axis located at the center of lane 1. The origin of the s-axis is located at the input node on the left with the direction of travel to the right. For the first 30m of the link vehicles are restricted from changing lanes, in order to prevent a vehicle from moving directly into the path of a vehicle that has not yet entered the network. As the distance to be traveled by each vehicle in the network is the same (5000m), we will substitute the term $L/T_d$ in \eqref{eq:vd} with $v_{do}$ the base desired velocity of a given vehicle. The base desired velocities of vehicles were distributed using the default speed distribution in VISSIM with a mean of approximately 87 km/hr. 

Simulations were run for 30 minutes of simulation time with CAV penetration rates of 0, 25, 50, 75 and 100\% and traffic demands from 2000 veh/hr up to 7500 veh/hr, increasing in 500 veh/hr increments. For all future discussions, the baseline scenario will be 0$\%$ CAV penetration at the respective traffic demand with ALL vehicles being HDVs. In post processing, as the network starts the simulation empty, we omit the data prior to the simulation time at which the network occupancy reaches 90\% of the maximum observed. We will refer to the remaining simulation time as the evaluation duration. 
HDVs are assumed to follow the Wiedemann-99 psycho-spacing car-following model \cite{Liu2016}, and the default rule-based lane selection (RBLS) algorithm of VISSIM, which was originally developed by Sparmann \cite{VISSIM}.

The Wiedemann Car-Following model assumes there are 5 driving states: 1)  "Free flow"; 2) Following; 3) Approaching; 4) Braking; and 5) Collision \cite{VISSIM}. The boundaries of these states are determined based on a variety of parameters including: a standstill distance $CC0$, a time headway $CC1$, and other parameters determining the oscillatory following behavior and free-flow acceleration levels \cite{VISSIM,Liu2016}. While VISSIM treats a given vehicle's (driver's) Wiedemann car following model as deterministic, the parameters are varied from vehicle to vehicle based on pre-defined probability distributions. Default parameters and distributions were utilized for all parameters except the standstill distance $CC0$ and time headway $CC1$, which were defined as $CC0 = 3.04m$ and $CC1 \sim \mathcal{N}\left(1.45 s, 0.01 s^2\right)$. As for the RBLS algorithm, default parameters were utilized, with free lane selection, where there are no rules/precedent that dictates a given lane being a fast/passing, travel, or slow lane.

Lastly, in the event the DMPC solver was unable to find a solution to the OCP (due to infeasible constraints), the CAV enters a fallback routine where the Wiedemann-99 car-following model and the RBLS algorithm are utilized for longitudinal and lateral control, respectively. In general such events are found to be rare and decrease with an increase in CAV penetration. Specifically, at low CAV penetration less than $2\%$ of all MPC solver calls do not find a solution in a given simulation and reducing to less than 0.02\% at 100\% CAV penetration.

\vspace{-0.1in}
\subsection{Benchmark 1-Dimensional Planner}

For further understanding of the impact of our 2D maneuver (speed and lane) planner, the proposed coordinated DMPC framework will also be compared to a 1D speed trajectory planner, intended to represent a state-of-the-art CACC. In order to isolate the benefits of planning laterally, we attempt to maintain the same underlying control framework for the 1D and 2D planners. To this end, we make a few modifications to the framework presented in Fig. \ref{fig:ctrl_framework}:
\vspace{-0.03in}
\begin{itemize}
\item only $a_d$ is passed from the DMPC block to the $c_i$ - motion block (i.e. the connection passing $\psi_d$ is removed, however, the DMPC formulation remains the same);
\item a human driver (HD) model block is added that utilizes the RBLS algorithm of VISSIM to evaluate $\psi_d$ and pass $\psi_d$ to the $c_i$ - motion block; and
\item a reference speed filter (described below) is added in between the RSA and DMPC blocks of Fig. \ref{fig:ctrl_framework}.
\end{itemize}
\vspace{-0.03in}
The reference speed filter works as follows: if the vehicle occupies lane $l$ at the current time, is predicted to be in lane $l$ by the DMPC at time step $k$ in the future, or lane $l$ is the left/right adjacent lane when a left/right turn signal from the HD block is received, then $v_{l,k}$ is unmodified from the RSA block; otherwise, if none of the prior mentioned conditions are met, $v_{l,k} = \epsilon$, where $\epsilon$ is a small positive number. In the following results and discussions we will refer to the 2D maneuver planning and 1D speed trajectory planning DMPC frameworks as the 2D planner and 1D planner, respectively.

\vspace{-0.16in}
\subsection{Results and Analysis}

We will now discuss the impacts of the proposed distributed maneuver planning framework on capacity and network flow, followed by the impacts on fuel economy at various CAV penetrations.

\subsubsection{Network Flow}

To evaluate the impact on traffic flow/throughput, we analyze the relationship between average travel speed to average traffic density. The average traffic density $\rho$ is calculated over the evaluation duration as follows \cite{Gazis2002}:
\begin{equation}
\rho = \frac{TTS}{LT_e},
\end{equation}
where $TTS$ is the total time spent on the network, and $T_e$ is the total time in the evaluation duration. Further, $TTS = \sum_{i=1}^{N_e}{T_i}$, where $N_e$ is the number of vehicles that were in the network over the evaluation duration, and $T_i$ is the travel time of the $i^{th}$ vehicle. The average velocity $\mu_v$ of traffic over the evaluation duration is then defined as $\mu_v = TDT/TTS$, where $TDT$  is the total distance traveled. We will define the total distance traveled as $TDT = \sum_{i=1}^{N_e}{D_i}$, where $D_i$ is the distance traveled by the $i^{th}$ vehicle.

\begin{figure}[t]
\parbox{3.5in}{\centering\includegraphics[width=3.2in]{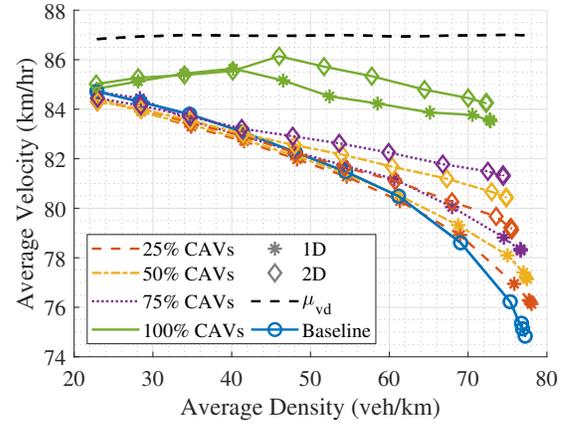}}
\caption{Velocity versus density diagram for both planners at varying penetration levels, where $\mu_{\text{vd}}$ is the average base desired velocity.}
\label{fig:vel_MFD} 
\end{figure} 
Fig. \ref{fig:vel_MFD} presents comparative plots of the 1D and 2D planners for varying levels of CAV penetrations. For a benchmark performance, we have included the average base desired velocity $\mu_{v_{do}}$ of the traffic. As the controller attempts to track the estimated average velocity of traffic, which is in turn dependent on the base desired velocity distribution, the (ideal) optimal average velocity of traffic is equivalent to $\mu_{v_{do}}$. We will group the average densities into 3 categories: low density ($\rho <$ 40 veh/km), moderate density (40 veh/km $\leq \rho \leq$ 60 veh/km), and high density ($\rho >$ 60 veh/km). At the tested CAV penetrations below 100\% and low densities, there is not a significant difference in average travel speed between the baseline (0\%) and the two planners. With partial CAV penetration, as the density increases to within the moderate range, only the 2D planner is able to outperform the baseline in terms of improving average traffic velocity or throughput. Yet at high densities, both planners are able to outperform the baseline (0\%). At 25\% to 75\% CAV penetration, both the 1D and 2D planners flatten the curve as traffic density increases (thereby achieving higher average travel velocities). However, at 100\% CAV penetration there is a significant change in the average velocity characteristics with respect to density. This change is attributable to the prediction of neigboring vehicles, where while lower CAV penetrations rely partially on trajectory predictions of neighboring HDVs that are subject to error, the 100\% CAV case uses solely communicated information that is more reliable. At 100\% CAV penetration and moderate to high densities, the 2D planner consistently achieves a higher average traffic velocity than the 1D planner. With both planners, at 100\% CAV penetration, there is an inflection point in Fig. \ref{fig:vel_MFD} around 40 or 46 veh/km for the 1D or 2D planner, respectively. As the density (or the number of communicating CAVs $N_p$) increases the difference between the lane reference speed calculated in  \eqref{eq:lane_ref_speed} and the actual average speed of traffic decreases. Thus, the inflection point marks where the traffic is sufficiently dense, such that the error in lane reference speed estimation is negligible.

In order to see how the improvements in average traffic velocity discussed above impact traffic flow, we define the average traffic flow rate $Q_a$ as \cite{Gazis2002}:
\begin{equation}
Q_a = \frac{TDT}{LT_e}.
\end{equation}
It should be noted that the observed average flow of traffic may not always meet demand, therefore we will differentiate between the prescribed traffic demand $Q_d$ and observed average traffic flow $Q_a$. 
\begin{figure}[t]
\parbox{3.5in}{\centering\includegraphics[width=3.3in]{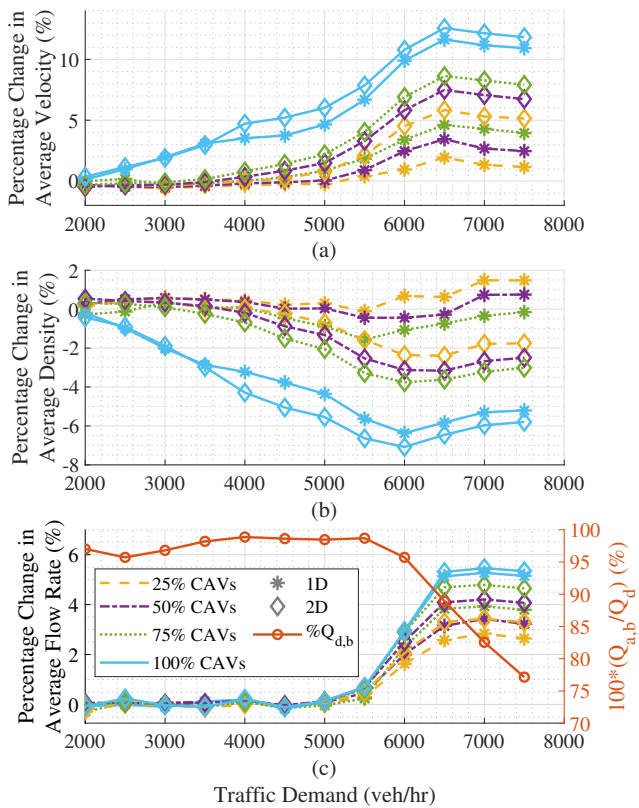}}
\caption{Percentage change in (a) average velocity, (b) average density, and (c) average flow rate relative to the baseline scenario for  25, 50, 75, and 100\% CAV penetration. $Q_{a,b}$ is the observed average flow rate for the baseline scenario. \vspace{-0.15in}}
\label{fig:percImproveFlow} 
\end{figure}
Fig. \ref{fig:percImproveFlow} presents the percentage change relative to the baseline for (a) average velocity, (b) average density, and (c) average flow rate with respect to traffic demand. The percentage change relative to the baseline is defined as $100\%[(scenario-baseline)/baseline]$, where $scenario$ and $baseline$ are the appropriate average metrics for the given simulation scenario and baseline, respectively. For reference and comparison to Fig. \ref{fig:vel_MFD}, a traffic demand of 2500 veh/hr resulted in densities around 28 veh/km, and demands of 4000 veh/hr resulted in densities between 45 and 50 veh/km depending on the CAV penetration. Despite improvements in average velocity being observed for select CAV penetration rates at traffic demands below 5500 veh/hr, due to a decrease in density, no improvement in overall vehicle flow rate is observed until a demand of over 5500 veh/hr. Referring to the right axis of Fig. \ref{fig:percImproveFlow}(c), where the baseline flow rate percent of demand $\%Q_{d,b}$ is presented, it can be noted that at traffic demands at or below 5500 veh/hr, the baseline meets the demand requirements. Above 5500 veh/hr, the baseline scenario no longer is able to meet demand requirements and both the planners are able to improve traffic flow by 2.5 to 5.5\% depending on CAV penetration. However, it should be noted, that these improvements in traffic flow are not sufficient to meet the demand. At demands over 6000 veh/hr, the 2D planner is able to obtain a higher average flow rate compared to the 1D planner.

\begin{figure}[t]
\parbox{3.5in}{\centering\includegraphics[width=3.55in]{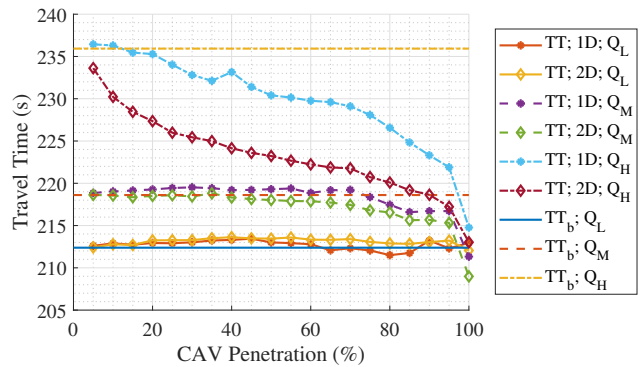}}
\caption{Average travel time versus penetration rate at traffic demands $Q_L = 2000$, $Q_M = 4000$, and $Q_H = 6000$ veh/hr. TT$_b$ is the average travel time of the baseline. \vspace{-0.15in}}
\label{fig:travel_time} 
\end{figure}

For the remainder of the results we will focus on three scenarios that correspond to low, medium and high traffic demands: $Q_L=2000$ veh/hr, $Q_M = 4000$ veh/hr, and $Q_H = 6000$ veh/hr, respectively. Furthermore, simulations with additional CAV penetration rates were completed between 0\% and 100\% at 5\% increments. Fig. \ref{fig:travel_time} presents the average travel times for the three traffic demand levels for planners as a function of CAV penetration. The observations parallel those made regarding the average velocity above and we omit the discussions for brevity.

\subsubsection{Fuel Economy}
 
Given the traffic micro-simulation outputs (speed trajectories), the fuel consumption of the relevant population was estimated using the methods and models presented in \cite{Dollar2017}. We define the percent reduction in fuel consumption FC\% as follows:
\begin{equation}
\begin{matrix}
\text{FC\%} = 100\left(1-\frac{\text{FC}}{\text{FC}_{b}}\right),
\end{matrix}
\end{equation}
where FC and $\text{FC}_b$ are the average fuel consumption rate for the given scenario and the associated baseline scenario, respectively. As the average velocity from scenario to scenario changes, we will also introduce the travel time adjusted percent reduction in fuel consumption AFC\%:
\begin{equation}
\text{AFC\%} = \text{FC\%}-100\left(\frac{\text{RFC}-\text{RFC}_{b}}{\text{FC}_{b}}\right).
\end{equation}
where RFC and $\text{RFC}_b$ are the fuel consumption rates required to maintain a constant velocity at the observed average velocity for the given scenario and the baseline scenario, respectively.

\begin{figure}[t]
\parbox{3.5in}{\centering\includegraphics[width=3.55in]{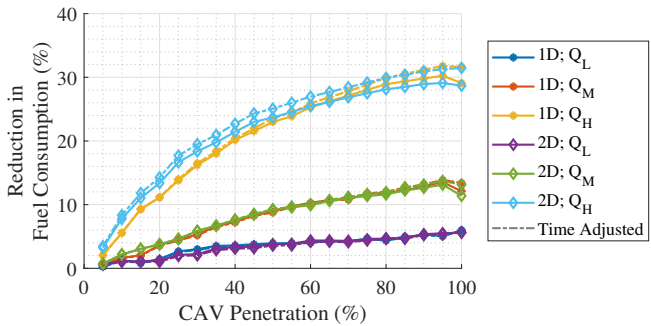}}
\caption{Percent reduction in fuel consumption versus penetration rate at traffic demands $Q_L=2000$, $Q_M=4000$, and $Q_H = 6000$ veh/hr.}
\label{fig:fuel_econ_percent} 
\end{figure}
Fig. \ref{fig:fuel_econ_percent} presents the FC\% and AFC\% for both planners as a function of CAV pentration at the three traffic demands $Q_L$, $Q_M$, and $Q_H$. All scenarios are able to meet or reduce fuel consumption compared to the baseline. The underlying reason for the fuel saving effect of the 2D planner has been investigated in our previous paper \cite{Goulet2020}. It is shown there that the 2D planner decreases variations in velocity for individual vehicles, making all vehicles travel with near constant velocity at close to the mean traffic speed. As the 1D planner utilizes the same logic for longitudinal control as the 2D planner, it realizes a similar effect. Specifically, both DMPC frameworks are able to reduce the sum of acceleration magnitudes over a given simulation from the baseline range of $\left(0.1,1.8\right)*10^6$ m/s\textsuperscript{2}, to a range of $\left(0.8, 1.2\right)*10^4$ m/s\textsuperscript{2} at $100\%$ CAV penetration. It has been shown in \cite{Dollar2017} that reductions in accelerations and decelerations lead to reduction in fuel consumption.

We seek to determine if the added degree of freedom in the lateral direction within the 2D planner contributes to the fuel savings. When comparing the two planning methods, in general, there is not a significant difference in the fuel consumption rate. The largest observed difference is at a traffic demand of $Q_H$ and 25\% CAV penetration, where the 2D planner improves FC\% by almost 3\% more than the 1D planner and AFC\% by almost 4\%. As penetration increases, at high traffic demand $Q_H$, the performance of the two planners converges. The reason the 2D planner does not improve FC\% and AFC\% significantly compared to the 1D planner, is that there is minimal room for improvement. In both cases, the goal of the controller is to track reference speeds while minimizing accelerations, therefore, based on the RSA, the optimal solution without disturbances would result in the RFC. 

\begin{figure}[t]
\parbox{3.5in}{\centering\includegraphics[width=3.55in]{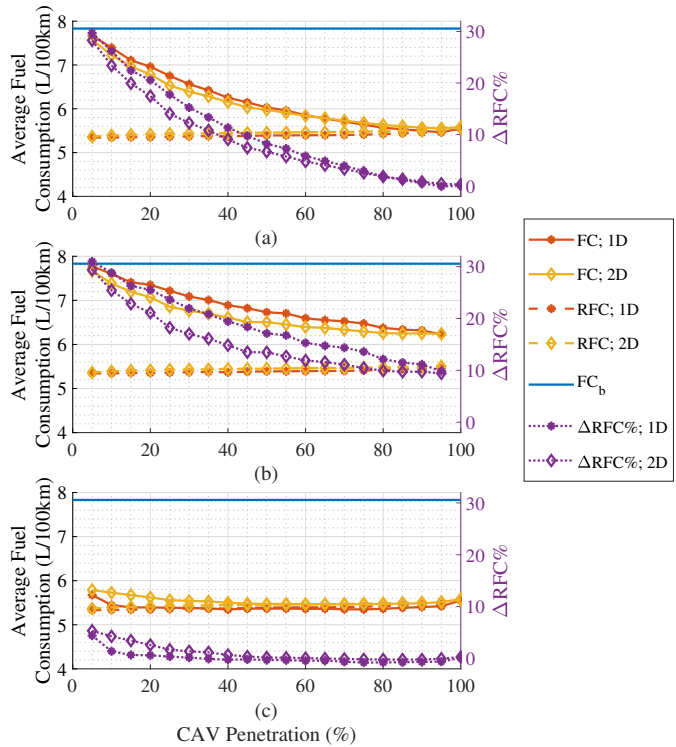}}
\caption{Fuel Economy results in (L/100km) versus penetration rate for the 1D and 2D planners at $Q_H=6000$ veh/hr traffic demands for (a) the entire mixed fleet, (b) HDVs only, and (c) CAVs only.}
\label{fig:fuel_econ} 
\end{figure}

To illustrate this, we define $\Delta RFC\%$,  the percentage difference of the observed fuel consumption rate to the required fuel consumption rate, as:
\begin{equation}
\Delta RFC\% = \frac{FC-RFC}{FC_b}.
\end{equation}
We present the fuel consumption rate and $\Delta RFC\%$ results for $Q_H$ as an example in Fig. \ref{fig:fuel_econ}.
The figure separates the average fuel consumption into different subsets of the vehicle population, specifically, the entire mixed fleet (both CAVs and HDVs) in Fig. \ref{fig:fuel_econ}(a), only the HDV population in Fig. \ref{fig:fuel_econ}(b), and only the CAV population in Fig. \ref{fig:fuel_econ}(c). When referencing the CAV population in Fig. \ref{fig:fuel_econ}(c), it can be seen that for both planners there is a negligible difference between RFC and FC above 50\% CAV penetration. At CAV penetrations below 50\%, the 1D planner actually results in marginally reduced FC for CAVs than the 2D planner. However, as seen in Fig. \ref{fig:fuel_econ}(b), the 2D planner results in a reduction in fuel consumption compared to the 1D planner for neighboring HDVs, which make up the majority of traffic. This results in a marginal reduction in fuel consumption of the entire mixed fleet (see Fig. \ref{fig:fuel_econ}(a)) for the 2D planner compared to the 1D planner at CAV penetrations below 50\%. 

\subsubsection{Lateral Maneuvers}
The difference between the 1D and 2D planners is their lane selection algorithms. Therefore, here we investigate the differences in how CAVs change lanes between the two cases. Further, we will investigate if this has an impact on how HDVs change lanes. 
\begin{figure}[t]
\parbox{3.5in}{\centering\includegraphics[width=3.55in]{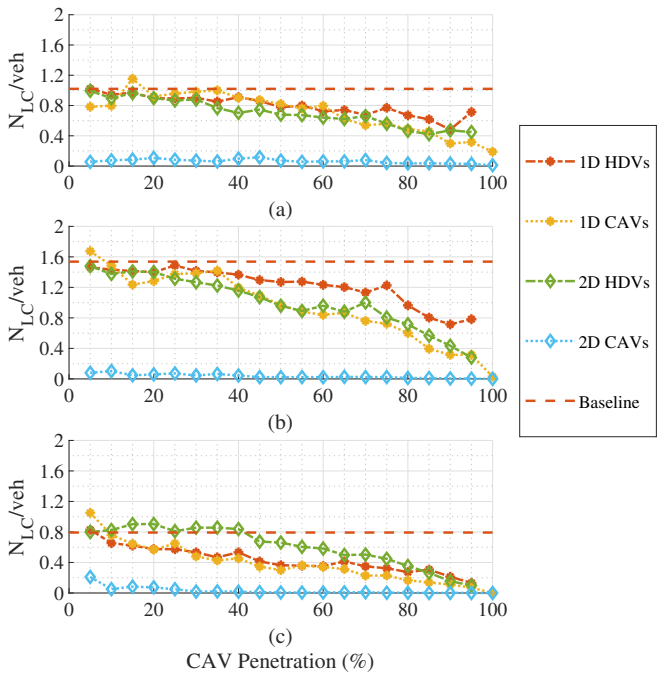}}
\caption{Average number of lane changes per vehicle versus CAV penetration rate at traffic demands (a) $Q_L$, (b) $Q_M$, and (c) $Q_H$.}
\label{fig:Nlc} 
\end{figure}
Fig. \ref{fig:Nlc} presents the average number of lane changes per vehicle, N$_\text{LC}$/veh, for both CAVs and HDVs for the three traffic demands and two control methods. In all scenarios, the 2D CAVs change lanes the least often, typically close to 0 N$_{\text{LC}}$/veh, as there is little incentives for 2D CAVs to change lanes. The reason being, the lane reference speeds are assigned based on the estimated average speed within the given lane, and each lane is subject to the same underlying distribution. The highest number of lane changes for the baseline scenario occurs at $Q_M$, in Fig. \ref{fig:Nlc}(b), as at lower demands there is less incentives to change lanes as traffic is sparse, and at higher demands there is less space for safe lane changes. In the case of the 1D planner, both CAVs and HDVs tend to change lanes at relatively the same rate for all CAV penetrations where HDVs are still present for the low and high traffic demands. This does not hold for the scenarios with a medium traffic demand and penetrations over 40\% 1D CAVs, where the HDVs change at a marginally higher frequency. In mixed traffic with either planner, HDVs tend to change lanes at an equal or lower frequency than the baseline scenario, as the incentives for human drivers to change lanes reduces as the CAVs smooth traffic to an average velocity. Further, regardless of the control framework used by CAVs, the N$_{\text{LC}}$/veh of HDVs tends to follow similar trends as CAV penetration increases.

\vspace{-0.12in}
\section{Conclusion}\label{sec:Conclusion}

In this paper, we proposed two versions of a distributed maneuver planning framework with provisions of sharing motion plans between connected vehicles and including a distributed reference speed assigner that fosters traffic speed harmonization. The framework is formulated to be scalable to operate with mixed traffic at different levels of CAV penetrations. Our evaluations of the framework in traffic micro-simulations indicate that both version of the framework can improve traffic throughput (by as much as 10\%) while offering a traffic fuel consumption reduction of as much as 30\%. We also observed that lane selection (whether rule-based or optimization-based) has the most impact at moderate traffic demands where frequent lane changes are feasible and likely, and that the speed harmonization reduces the incentive to change lanes for all vehicles (including HDVs) as the CAV penetration increases.

Directions for future work in this area include refining the lane reference speed assigner by removing some of the assumptions made here, and also investigating computationally tractable interaction-aware object vehicle state prediction approaches. In addition, approaches for incorporating communal traffic efficiency goals in similar distributed implementations are of interest.

\appendices

\vspace{-0.15in}
\section{Augmented State Dynamics of CAV $c_i$}\label{sec:aug_state}

The augmented state of the ego CAV is presented in \eqref{eq:aug_ego_state}, which includes the particle dynamics model in the Frenet frame as presented in \eqref{eq:ego_model} and the additional states introduced for computational modeling purposes throughout Section \ref{sec:ControlFramework}.
\begin{equation}\label{eq:aug_ego_state}
\dot{x} = \begin{bmatrix}\dot{s}\\
\dot{y}_e\\
\dot{v}_t\\
\dot{\psi}\\
\dot{a}_t\\
\dot{\psi_d}\\
\dot{\zeta}\\
\dot{d_1}\\
\vdots\\
\dot{d}_{N_l-1}\end{bmatrix} = \begin{bmatrix}\frac{v_t}{1-y_e\kappa\left(s\right)}\cos{\psi}\\
v_t\sin{\psi}\\
a_t\\
\tau_{\psi}\left(\psi_d-\psi\right)\\
\tau_a\left(a_d-a_t\right)\\
\delta\psi_d\\
u_{\zeta}\\
u_{d_1}\\
\vdots\\
u_{d_{N_l-1}}\end{bmatrix}.
\end{equation}

\vspace{-0.1in}
\section{Hyperellipse Half Minor and Major Axes Definition}\label{sec:ov_ellipse}

\begin{figure}[t]
\parbox{3.375in}{\centering\includegraphics[width=3in]{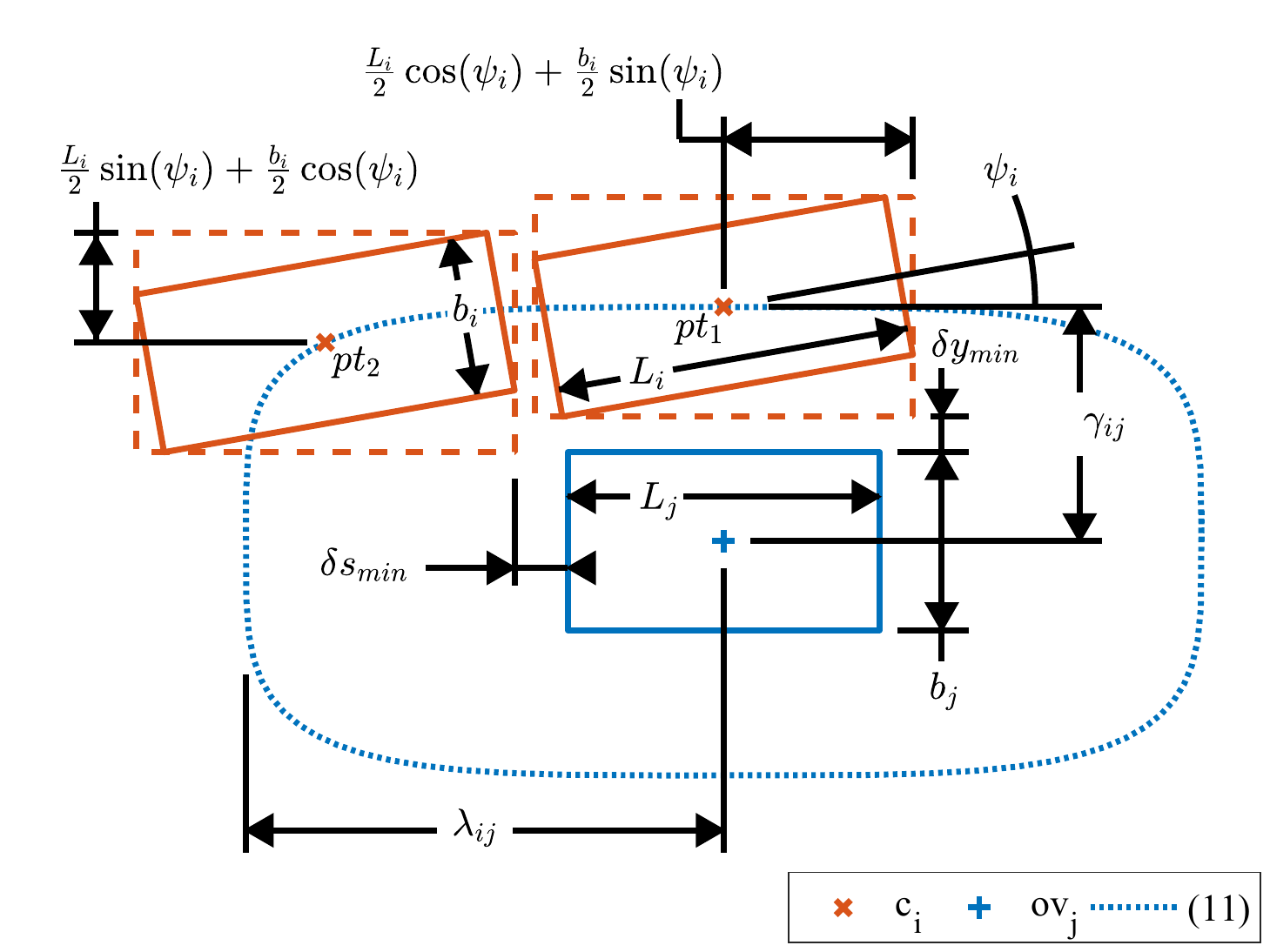}}
\caption{Schematic depicting derivation of the axes of the OV hyperellipse. \vspace{-0.2in}}
\label{fig:ellipse} 
\end{figure}

We define the hyperelliptical constraint using two positions of CAV $c_i$ relative to $ov_j$, see Fig. \ref{fig:ellipse} for a schematic. 
The first, $pt_1$, ensures that a minimum lateral distance $\delta y_{min}$ is maintained between CAV $c_i$ and $ov_j$. The second, $pt_2$, is chosen to ensure the right front corner of CAV $c_i$ will not contact the left rear corner of $ov_j$ along with an additional safety distance in the $s$-direction $\delta s_{min}$. The half minor and 
major axes can then be computed as follows:
\begin{subequations}
\begin{gather}
\gamma_{ij} = \frac{L_i}{2}\sin\left(\psi_i\right) + \frac{b_i}{2}\cos\left(\psi_i\right) + \frac{b_j}{2} + \delta y_{min}\\
\lambda_{ij} = \frac{\frac{L_i}{2}\cos\left(\psi_i\right) + \frac{b_i}{2}\sin\left(\psi_i\right) + \frac{L_j}{2} + \delta s_{min}}{\left[1 - \left(\frac{\gamma_{ij} - \delta y_{min}}{\gamma_{ij}}\right)^4\right]^{1/4}},
\end{gather}
\end{subequations}
where $b_i$ and $L_i$ are the width and length of CAV $c_i$, respectively.

\vspace{-0.1in}
\section{Algorithm for Calculating Unique Field of View}\label{app:unique_FOV}
The algorithm used to perform a pairwise comparison of FOV bounds and calculate the unique area $A_{u,lp}$ of CAV $c_p$'s FOV is presented in Algorithm \ref{algo1}. We define the set of all FOV bounds of CAV $c_p$ in lane $l$ as $\mathbf{FOV}_{lp} = \begin{Bmatrix}FOV_{1,lp}&...&FOV_{N_{FOV,p},lp}\end{Bmatrix}$, where $N_{FOV,p}$ is the number of FOV bound subsets. Further the FOV bound subset $FOV_{q,lp}=\begin{Bmatrix}\lowerb{\alpha}_{q,lp},&\upperb{\alpha}_{q,lp}\end{Bmatrix}$ and the width of lane $l$ is $w_l$.
\begin{algorithm}
\caption{CAV $c_i$ Calculating $A_{u,lp}$}
\label{algo1}
\begin{algorithmic}[1]\label{algo:Aulp}
\For{each lane $l \in \mathcal{L}$}
	\State Initialize $FOV_{0,li} = \begin{Bmatrix}\lowerb{\alpha}_{li},& \upperb{\alpha}_{li}\end{Bmatrix}$
	\For{each communicating CAV $c_p \in \mathcal{C}_i$ in lane $l$}
		\State Initialize $FOV_{0,lp} = \begin{Bmatrix}\lowerb{\alpha}_{lp},& \upperb{\alpha}_{lp}\end{Bmatrix}$
		\State Initialize flag $accounted_{lp} = False$
		\For{each set of bounds $FOV_{q,li}$}
			\For{each set of bounds $FOV_{r,lp}$}
				\If{$FOV_{r,lp}$ encompasses $FOV_{q,li}$}
					\State $FOV_{q,li} =  FOV_{r,lp}$
					\State split $FOV_{r,lp}$
					\State $N_{FOV,p} = N_{FOV,p}+1$
					\State $accounted_{lp} = True$
				\ElsIf{$FOV_{r,lp}$ overlaps $FOV_{q,li}$}
					\State shrink $FOV_{r,lp}$ bounds appropriately
					\State enlarge $FOV_{q,li}$ bounds appropriately
					\State $accounted_{lp} = True$
				\ElsIf{$FOV_{q,li}$ encompasses $FOV_{q,lp}$}
					\State  $FOV_{r,lp} = \begin{Bmatrix}0,&0\end{Bmatrix}$
					\State $accounted_{lp} = True$
				\ElsIf{$FOV_{r,lp}$ is completely unique}
					\If{$accounted_{lp}=False$}
						\State $N_{FOV,i} = N_{FOV,i}+1$
						\State $FOV_{N_{FOV,i},li} = \begin{Bmatrix}\lowerb{\alpha}_{r,lp},&\upperb{\alpha}_{r,lp}\end{Bmatrix}$
						\State $accounted_{lp} = True$
					\EndIf
				\EndIf
			\EndFor
			\For{each set of bounds $FOV_{m,li}$ ($m \neq q$)}
				\If{one FOV encompasses the other}
					\State remove the encompassed FOV
				\ElsIf{$FOV_{m,li}$ overlaps $FOV_{q,li}$}
					\State enlarge the FOV with the smaller index
					\State remove the FOV with the larger index
				\EndIf
			\EndFor
		\EndFor
		\State $A_{u,lp} = w_l\sum_{r=1}^{N_{FOV,p}}\upperb{\alpha}_{r,lp}-\lowerb{\alpha}_{r,lp}$
	\EndFor
\EndFor
\end{algorithmic}
\end{algorithm}

\section*{Acknowledgment}

The authors acknowledge financial support provided by the US Department of Energy under contract No. DE-EE0008232 for this research. The opinions and results expressed in this work are solely the reprehensibility of the authors.

\ifCLASSOPTIONcaptionsoff
  \newpage
\fi



\bibliographystyle{IEEEtran}
\bibliography{IEEEabrv,T-ITS-20-05-0996_final}
%
%
%

%

\begin{IEEEbiography}[{\includegraphics[width=1in,height=1.25in,clip,keepaspectratio]{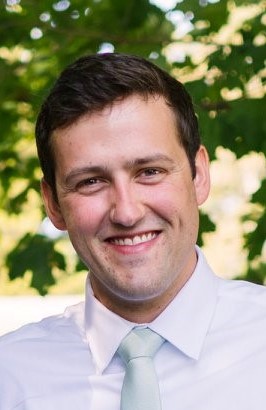}}]{Nathan Goulet}
received his BS (2014) degree in Mechanical Engineering, as well as, his BA (2014) in German Studies from the University of Connecticut, before working in the Automotive industry as a design engineer. Since 2017, he has been in pursuit of his Ph.D. degree in Automotive Engineering at the Clemson University-International Center for Automotive Research. His research interests include optimal control, distributed control of automated vehicles, and vehicle state estimation and trajectory prediction.
\end{IEEEbiography}

\begin{IEEEbiography}[{\includegraphics[width=1in,height=1.25in,clip,keepaspectratio]{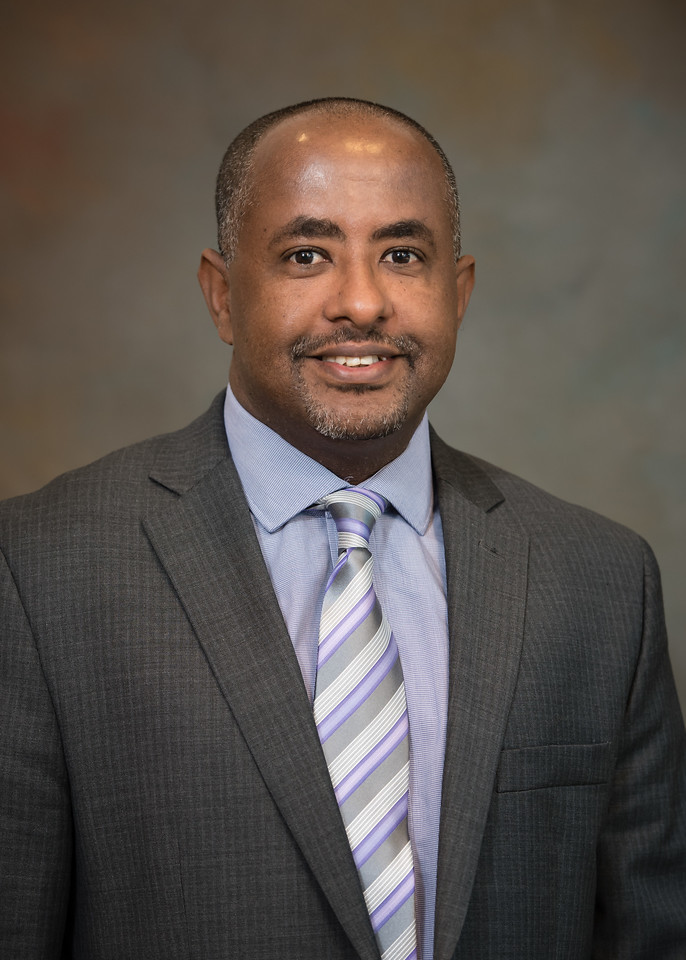}}]{Beshah Ayalew}
is a Dean’s Distinguished Professor of Automotive Engineering at the Clemson University-International Center for Automotive Research. He received his MS (2000) and Ph.D. (2005) degrees in Mechanical Engineering from Penn State University. His interest and expertise are in controls and dynamical systems with applications in connected and automated vehicle traffic systems and energy systems. Dr. Ayalew has received the Ralph R. Teetor Educational Award from the Society of Automotive Engineers (SAE) International (2014), the Clemson University Board of Trustees Award for Faculty Excellence (2012, 2019), and the National Science Foundation’s CAREER Award (2011). He was also a recipient of the Penn State Alumni Association Dissertation Award (2005). He is a senior member of IEEE, a fellow of ASME, and a member of SAE.  Dr. Ayalew has authored/co-authored more than 150 refereed publications.  He currently serves as an associate editor for IEEE’s Transactions on Intelligent Transportation Systems.
\end{IEEEbiography}







\end{document}